\documentclass[twocolumn,prx,twocolumn,superscriptaddress]{revtex4-1}
\usepackage{amsmath,amssymb,mathrsfs}
\usepackage{natbib}
\usepackage{subfigure}
\usepackage{tabularx}
\usepackage{epsfig}
\usepackage{longtable}
\usepackage{amsfonts}
\usepackage{rotating}
\usepackage{subfigure}
\usepackage{amsmath}
\usepackage{comment}
\usepackage{bbold} 
\usepackage{hhline}
\usepackage{braket}
\usepackage{txfonts, comment}

\def\be{\begin{equation}}
\def\ee{\end{equation}}
\def\bea{\begin{eqnarray}}
\def\eea{\end{eqnarray}}
\def\nn{\nonumber}

\def\up{\uparrow} 
\def\down{\downarrow}

\usepackage[unicode=true,bookmarks=true,bookmarksnumbered=false,bookmarksopen=false,breaklinks=false,pdfborder={0 0 1},backref=false,colorlinks=true]{hyperref}

\hypersetup{linkcolor=magenta,urlcolor=blue,citecolor=blue,pdfstartview={FitH},hyperfootnotes=false,unicode=true}

\def\be{\begin{equation}}
\def\ee{\end{equation}}
\def\bea{\begin{eqnarray}}
\def\eea{\end{eqnarray}}
\def\nn{\nonumber}

\def\up{\uparrow}
\def\down{\downarrow}

\begin{document}
\title{Programmable Quantum Annealing Architectures with  Ising Quantum Wires} 
\author{Xingze Qiu} 
\affiliation{State Key Laboratory of Surface Physics, Institute of Nanoelectronics and Quantum Computing, and Department of Physics, Fudan University, Shanghai 200433, China}
\author{Peter Zoller} 
\affiliation{Center for Quantum Physics, University of Innsbruck, 6020 Innsbruck, Austria}
\affiliation{Institute for Quantum Optics and Quantum Information of the Austrian Academy of Sciences, 6020 Innsbruck, Austria} 

\author{Xiaopeng Li} 
\email{xiaopeng\_li@fudan.edu.cn}
\affiliation{State Key Laboratory of Surface Physics, Institute of Nanoelectronics and Quantum Computing, and Department of Physics, Fudan University, Shanghai 200433, China}
\affiliation{Shanghai Qi Zhi Institute, AI Tower, Xuhui District, Shanghai 200232, China} 

\begin{abstract}

Quantum annealing aims at solving optimization problems efficiently by preparing the ground state of an  Ising spin-Hamiltonian quantum mechanically.  A prerequisite of building a quantum annealer is the implementation of programmable long-range two-, three- or multi-spin Ising interactions. We discuss an architecture, where the required spin interactions are implemented via two-port, or in general multi-port quantum Ising wires connecting the spins of interest. This quantum annealing architecture of spins connected by Ising quantum wires can be realized by exploiting the three dimensional (3D) character of atomic platforms, including atoms in optical lattices and Rydberg tweezer arrays. The realization only requires engineering  on-site terms and two-body interactions between nearest neighboring qubits.  The locally coupled spin model on a 3D cubic lattice is sufficient to effectively produce arbitrary all-to-all coupled Ising Hamiltonians.  We illustrate the approach for few spin devices solving Max-Cut and prime factorization problems, and discuss the potential scaling to large atom based systems. 
\end{abstract}

\date{\today}

\maketitle

\section{Introduction}

There has been growing research interests in quantum annealing in the effort to speed up complex search and optimization problems~\cite{farhi_quantum_2000,2008_Chakrabarti_RMP,2018_Lidar_RMP}  including 
BQP, NP-complete and NP-hard problems~\cite{farhi_quantum_2000,2014_Lucas_FIP}. While a quantum annealer might not reduce the classical computation complexity of NP-hard problems 
$O(\exp(\alpha N^\gamma) )$ [with $N$ the problem-size] to polynomial~\cite{2008_Maggs_PRL,2011_Dickson_PRL,Altshuler12446,2014_Ruben_PRX,Ronnow_2014}, an exponential speedup for BQP has been suggested~\cite{2008_Du_Factorization,Kais_2018}, and one might gain significant improvement on coefficients $\alpha$ and $\gamma$ for NP problems~\cite{farhi_quantum_2000,2014_Lucas_FIP,zhou2018quantum} compared to classical algorithms.  
Because of important implication for both science~\cite{2014_Lucas_FIP} and commercial applications~\cite{king2017quantum}, quantum annealing has received significant attention in recent years~\cite{2014_Troyer_NPhys,Lechnere1500838,2016_Rocchetto_SA,2017_Glaetzle_NC,pichler2018computational,zhou2018quantum}. 

Among  various platforms considered for building a quantum annealer ~\cite{altman2019quantum,alexeev2019quantum},  cold atoms trapped in optical potential provide a scalable quantum simulation platform with  versatile controllability~\cite{gross2017quantum}, as demonstrated in experiments  emulating High-Tc superconductivity, quantum phase transitions and criticality, and quantum thermalization with atomic Hubbard models with tens to thousands of atoms~\cite{gross2017quantum}. Recent experiments have achieved single site control, and free programmability in optical lattices
~\cite{Yi_2008,gauthier2016direct,mazurenko2017cold,2019_Chin_PRX,2019_Porto_PRX,qiu2020precise}, and similar optical tweezer arrays provide us with large-spacing optical lattice with engineered spin-spin interactions through Rydberg dressing~\cite{2010_Buchler_PRL,2010_Pupillo_PRL,2011_Arimondo_PRL,2016_Weiss_Science,browaeys2020many}. Thus present atomic setups of engineered many-body systems provide us with new opportunities in building a quantum annealer.

 \begin{figure*}
\begin{center}
\includegraphics[width=.9\linewidth]{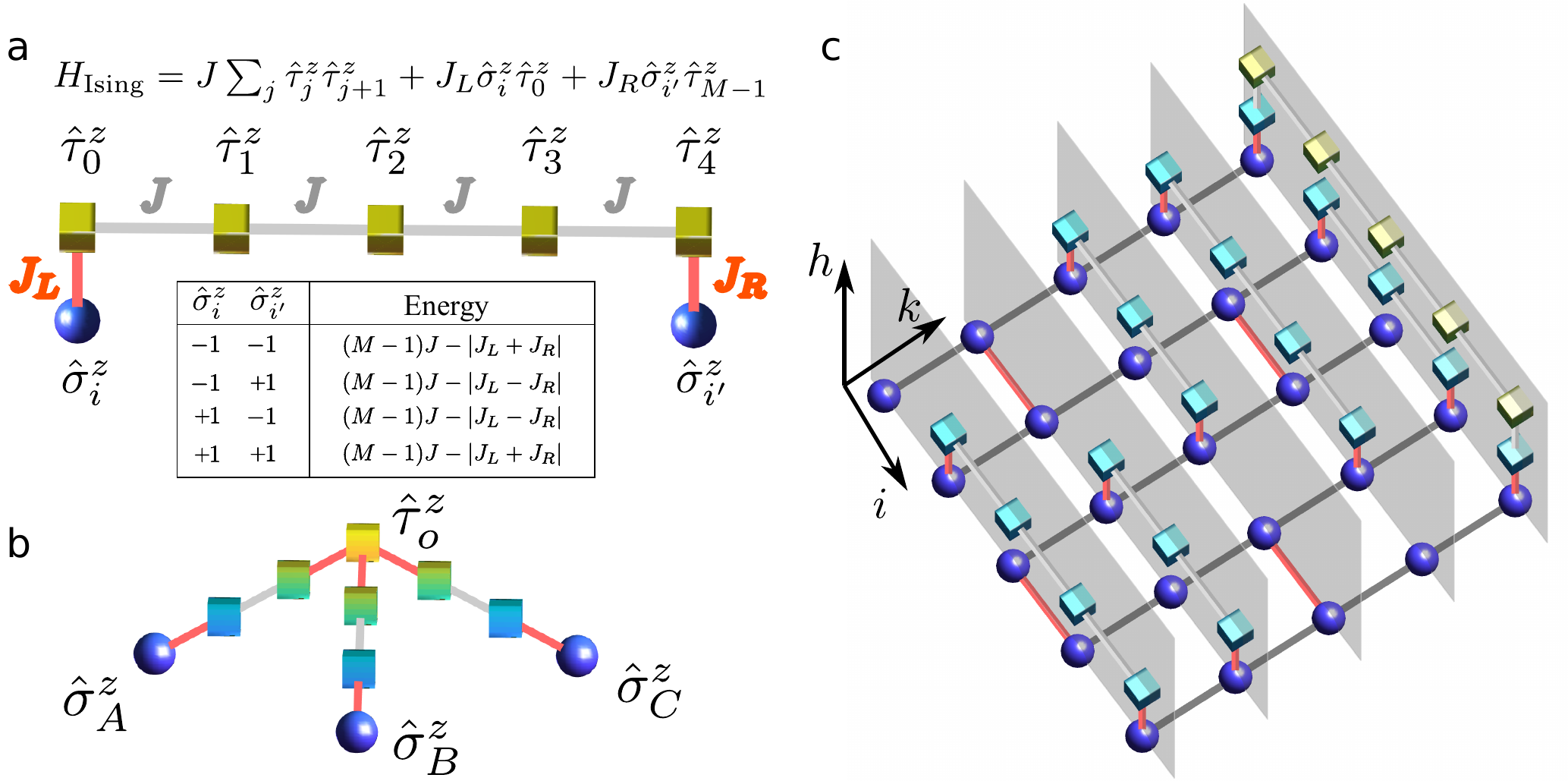}
\vspace{-2mm}
\end{center} 
\caption{{\bf Illustration of the three-dimensional cubic architecture for quantum annealing.} 
({\bf a}), the Ising ferromagnetic quantum wire induced interaction. With a ferromagnetic quantum wire coupling to two qubits $\hat{\sigma}^z_i$ and $\hat{\sigma}^z_{i'}$ at the two ends. The auxiliary spins in the wire are denoted as $\hat{\tau}^z_{j=0, 1, \ldots, M-1}$ (the number of ancilla $M$ is five in this example), which are ferromagnetically coupled by $J<0$, with Hamiltonian $H_{\rm Ising}$. The coupling between $\hat{\tau}^z_0$ ($\hat{\tau} ^z_{M-1}$) and $\hat{\sigma}^z _i$ ($\hat{\sigma}^z_{i'}$) is $J_L$ ($J_R$). 
The ground state energy of the quantum wire with the two distant qubits restricted to  an eigenstate subspace of $\hat{\sigma}^z_i$ and $\hat{\sigma}^z_{i'}$  
 is listed in the table, which determines the induced effective interaction between the two  qubits. ({\bf b}), the Y-junction connector for engineering a three-body interaction. With three quantum wires forming a Y-junction that couples $\hat{\sigma}_{A,B,C}^z$ through $\hat{\tau}_o^z$ (Eq.~\eqref{eq:HY}), an effective three-body interaction is mediated (see main text). 
 ({\bf c}), the geometry of the 3D cubic quantum annealing architecture.  
The blue spheres represent the $N$ duplicated copies of logical qubits. The cubes represent the introduced ancilla.  The light and dark grey links correspond to ferromagnetic couplings with fixed strengths $J$ and $J_d$. The red links correspond to the programmable Ising couplings (Eq.~\eqref{eq:Code}). All couplings in this quantum annealing architecture are local and at most between two  neighboring sites, and can be embedded  in a regular cubic lattice.  In this example we choose $N = 5$ for an illustration. 
}
\label{fig:f1}
\end{figure*}

However, in mapping optimization problems to a quantum annealer, all-to-all long-ranged couplings generically arise, which, for example, in atomic Hubbard models with optical lattices cannot be directly engineered. One way to approach the problem of long-range  interactions is to develop spin encodings in an enlarged spin space, where programmable long-range interactions are mapped to the experimentally simpler problem controlled local fields~\cite{Lechnere1500838}. This comes with the requirement, however, of implementing local four-body constraints, which can be resolved in terms two-body interactions as discussed in ~\cite{2017_Glaetzle_NC}. 
In contrast, we discuss below a completely distinctive scheme---a three dimensional 
(3D) architecture for quantum annealing (Fig.~\ref{fig:f1}), distant spins (or qubits) are coupled through an Ising ferromagnetic quantum wire playing the role of a connector mediating the interactions.  
The ancilla spins introduced as  quantum wires are carefully organized such that the 
3D quantum annealing architecture 
can be embedded into a regular cubic lattice with nearest neighboring interaction only. This geometric arrangement comes with the requirement to duplicate the Ising spins into multiple copies to provide ports for coupling to the wires, as illustrated in Fig.~\ref{fig:f1}. As discussed below, the 3D architecture  can be implemented by ground state atoms in optical lattices considering superexchange interactions, or the Rydberg-dressing induced couplings. The latter enjoys an advantage of having larger coupling strength.  
The scheme of quantum wire encoding programmable long-range interactions  is rather flexible and the  geometry can be adapted according to physical resources in actual experimental realization. 
We find reasonable scalability even in presence of thermally activated errors, potentially existent in experiments performing large-scale quantum annealing. 
This work  indicates  large-scale atom-based quantum annealing is accessible with near-term technology.  
We also provide a  generalization of the quantum wire construction to a quantum network mediating  $m$-body interactions in general.  
Experimental demonstration of the quantum wire mediating programmable $m$-body interactions does not only provide a building block of the 3D quantum annealing architecture, but also opens up novel opportunities for quantum engineering of exotic spin models and interesting quantum many-body physics.

\medskip 
\section{Quantum Hamiltonian Construction} 
\label{sec:HamConstruction} 
In annealing solving an optimization problem is formulated as finding the ground state of an Ising spin glass model, defined on an undirected graph $G = (V, E)$~\cite{2014_Lucas_FIP}. The corresponding classical Hamiltonian is
\be 
H [\{s\}]= \sum_{i=0} ^{ N-1} b_i s_i + \sum_{(ii')\in E} K_{ii'} s_i s_{i'} , 
\label{eq:IsingHam} 
\ee 
with $s_i$ an Ising variable ($\pm$), $i$ labeling the vertices of the graph, and $N$ the number of vertices. The local field $b_i$ and the Ising coupling $K_{ii'}$ are assumed programmable and encode the optimization problem to be solved --- for example, they depend  on graph-edges in graph partitioning~\cite{2014_Lucas_FIP}. 

In quantum annealing, the Ising variables are promoted to Pauli-z  operators ($\hat{\sigma}^z$) acting on qubits. Here the ground state is reached  following an adiabatic evolution of a time-dependent Hamiltonian of a quantum system~\cite{2018_Lidar_RMP}, 
\bea 
\label{eq:qIsingHam}
\hat{H}_{\rm QA} (\tau) &=&-\left[1-(\tau/\tau_{\rm ad}) \right] \sum_i \hat{\sigma}^x _i/2 \nn   \\ 
&+& (\tau/\tau_{\rm ad} ) 
	\left[ \sum_{i=0} ^{N-1} b_i \hat{\sigma}^z_i  + \sum_{(ii')\in E} K_{ii'} \hat{\sigma}^z _i \hat{\sigma}^z _{i'} \right] ,
\eea  
with the time $\tau \in [0, \tau_{\rm ad}]$, and $\tau_{\rm ad}$ the total evolution time. 
A challenge in implementing quantum annealing arises from the requirement to physically represent programmable infinite-range interactions represented by the matrix $K_{ii'}$, while physical resources available on quantum platforms are typically restricted to (quasi-)local two body couplings. While platforms like trapped ions~\cite{2010_Duan_RMP} and Rydberg arrays~\cite{2010_Saffman_RMP} provide long range interactions, the requirement of long-range interactions typcially interferes with scalability of the system. 
Below we address this problem by introducing quantum wires to connect physical qubits in a 3D geometry, as is often available in atomic platforms.

\medskip 
\subsection{Local Hamiltonian construction} 
{
Our construction starts by rewriting the interaction between the Ising variables $s_i$ in Eq.~\eqref{eq:IsingHam}, and similar in the quantum case with $s_i\rightarrow \hat \sigma^z_i$ in Eq.~\eqref{eq:qIsingHam}, as 
\bea  
\label{eq:domainwallHam} 
&& \textstyle K_{ii'} s_i s _{i'}  = 
{\rm min}_{n_{ii'} }\left\{  - 2J \sum_{(ii')\in E} n_{ii'}  \right.  \nn \\ 
&&\textstyle  \left. -\sum_{(ii')\in E}  2K_{ii'} [ n_{ii'} {\rm mod } \,2 - 1/2] s_i  s _{i'} \right\} , 
\eea
where $n_{ii'}$ is an auxiliary integer-valued degrees of freedom $n_{ii'} \ge 0$, and we  require the energy penalty $|J|$ dominates over the maximal coupling ($K_{\rm max} \equiv {\rm max}\{ |K_{ii'}| \}$) having, 
\be 
J  <- K_{\rm max}. 
\ee    
Although the coupling in Eq.~\eqref{eq:domainwallHam} still appears non-local, it can be realized by a quantum wire connecting Ising spins $i$ and ${i'}$. We choose an Ising ferromagnetic spin chain for the wire, and the auxiliary degrees of freedom $n_{ii'}$ corresponds to the number of domain wall defects in the spin chain  [Fig.~\ref{fig:f1}({\bf a})]. 
The mediated Ising interaction between $\hat{\sigma}^z_i$ and $\hat{\sigma}^z_{i'}$ is 
$(|J_{L,ii'}  - J_{R,ii'} | - |J_{L,ii'} + J_{R,ii'}| )/2 = -{\rm sgn}(J_{L,ii'} J_{R,ii'}) {\rm min} (|J_{L,ii'}|, |J_{R,ii'}|) $, 
 with $J_{L, ii'}$ ($J_{R,ii'}$) the coupling between $\hat{\tau}^z_i$ ($\hat{\tau}^z_{i'}$) and the leftmost (rightmost) ancilla [Fig.~\ref{fig:f1}({\bf a})]. 
We then set 
\be 
J_{L, ii'} = K_{ii'} , \,\,\, J_{R,ii'} = -|K_{ii'}|. 
\label{eq:JLR} 
\ee 
To the extent such quantum wires can be implemented in a physical platform, this achieves the required scalable and programmable long-range couplings. 
The above construction  is reminiscent of gauge fields mediating long-range interactions in field theories---for example long-range Coulomb interactions are mediated by fluctuating electromagnetic waves in quantum electrodynamics~\cite{weinberg_1995}. In general, the quantum wires connecting the two distant qubits can also be implemented with other spin  models or possibly even bosons, allowing the interaction mediated by the quantum wire to be analytically calculated. Here we choose the Ising ferromagnetic spin chain for the quantum wire for simplicity. In this construction, different quantum wires connecting a pair $(ii')$, are assumed  decoupled.  This poses a physical requirement on assembling these quantum wires without physical crossings.

Lining up the logical qubits $\hat{\sigma}^z_i$ from $i=0$ to $N-1$ in space, one problem arises that  it is fundamentally impossible to allocate these ancilla on a regular two dimensional lattice with a linear size of $N$, for  the total number of quantum wires scales as $N(N-1)/2$ and their lengths, $M_{ii'}$, are at the order of $N$  on average---$M_{ii'}\sim d_{ii'} \equiv |i-i'|\sim N $. 
To resolve this problem, we duplicate each logical qubit, $\vec{\hat{\sigma}}_i$ (Pauli operators) into $N$ copies, as $\vec{\hat{\sigma}}_{ik}$ with $k \in [0, N-1]$. 
 These duplicated qubits are ferromagnetically coupled through 
\be 
\hat{H}_{{\rm D}, i} = J_d \sum_{k= 0} ^{N-2 } \hat{\sigma}^z _{i,k} \hat{\sigma}^z _{i,k+1}, 
\label{eq:HamHd} 
\ee  
with the coupling strength $J_d<0$.

The qubit duplication allows us to assemble quantum wires in a three dimensional cubic lattice without any crossing. 
Fig.~\ref{fig:f1}({\bf c}) shows one example having $N = 5$.  
The duplicated logical qubits $\vec{\hat{\sigma}}_{ik}$ are placed at $(i, k, h=0)$. 
The ancilla quantum wire connecting qubits $i$ and $i'$, is placed in a two dimensional vertical layer with $k = \underline{ i+i'}$, with 
$\underline{i+i'}$ a shorthand notation for the modular summation ($i+i'$ mod $N$). 
The quantum wire connector Hamiltonian reads as 
\bea 
 \label{eq:IsingSpinChain} 
 \hat{H}_{{\rm QWC},ii'}  &=&   J \sum_{j=0} ^{M_{ii'} -2}  \hat{\tau}^z _{ii',j}  \hat{\tau}^ z_{ii', {j+1}} \nn  \\ 
&+&  \left[ K_{ii'}  \hat{\sigma}^z _{i,\underline{i+i'} }   \hat{\tau}^z _{ii', 0}  
-|K_{ii'}| \hat{\sigma}^z _{i', \underline{i+i'} }  \hat{\tau} ^z _{ii', M_{ii'}-1 } \right],   
\eea  
The quantum wire 
 starts from the position $(i, k = \underline{i+i'}, h=1)$, extends vertically first, and bends towards the $i$-axis 
at $(i,  \underline{i+i'} , h = \lfloor d_{ii'} /2 \rfloor)$, then bends downward at $(i', \underline{i+i'},  \lfloor d_{ii'} /2 \rfloor)$, 
 and reaches the end at $(i', \underline{i+i'},  1)$. 
The length of the ancilla quantum wire is 
$M_{ii'} = d_{ii'} + 2 \lfloor d_{ii'} /2 \rfloor -1 $, accordingly.  
The quantum wires having the same $\underline{i+i'}$ are placed in the same $k$-layer with $k =\underline{ i+i'}$ (see Fig.~\ref{fig:f1}).  
 The maximal height of the 3D quantum annealing architecture  along the $h$ direction is $h_{\rm max} = \left \lceil{N/2}\right \rceil -1$.

With the above construction, we reach an effective local quantum annealing  Hamiltonian for universal Ising spin glass problems given as, 
\bea 
\label{eq:Code} 
&& \hat{H} _{\rm LQA} (\tau) = 
[1-(\tau /\tau_{\rm ad}) ] \hat{H}_0 + (\tau /\tau_{\rm ad})  \hat{H}_{\rm P}, \nn \\  
&& \hat{H}_0 = - \sum_{ik} \hat{\sigma}^x_{ik} /2 - \sum_{(ii')\in E} \sum _{j = 0} ^{ M_{ii'} -1 }  \hat{\tau}^x _ {ii', j} /2, \\ 
&& \hat{H}_{\rm P} = \sum_{i=0}^{N-1} \left[ \hat{H}_{{\rm D}, i}+ b_i \hat{\sigma}^z _{i0}\right] 
+ \sum_{(ii')\in E} \hat{H}_{{\rm QWC}, ii'}. \nn
\eea 
 Here the local fields $b_i$ are applied to $\hat{\sigma}_{ik}^z$ with $k=0$ only, and an alternative way is to distribute that to all duplicated $N$ copies. 
 The ferromagnetic interactions, $J$ and $J_d$ that couple the ancilla and the duplicated logical qubits have fixed strengths, i.e., independent of the indices $i$ and $k$. 
The interactions between the ancilla and the duplicated logical qubits, $K_{ii'}$, encode the distant couplings in the Ising spin glass, and are required to be programmable.  
We note here that the quantum wire connector Hamiltonian,  $H_{{\rm QWC}, ii'}$, would reduce to a direct coupling 
$K_{ii'} \hat{\sigma}^z _{i, \underline{i+i'}}  \hat{\sigma}^z _{i', \underline{i+i'}}$ for $i' = i \pm 1$. 
In this way, we have constructed a local 3D quantum annealer for the all-to-all coupled spin glass which is embedded in a regular 3D cubic array of qubits with nearest neighbor interaction only. 
The size of this cubic lattice is $N \times N \times \left \lceil{N/2}\right \rceil$. The non-crossing requirement to avoid potential engineering difficulty in experiments is satisfied. 
}

We remark here that given the geometrical structure of the proposed 3D cubic architecture [Fig.~\ref{fig:f1}({\bf c})], it often permits efficient compression. 
For example, the vertical layers not containing any quantum wires can be removed, and the height of one quantum wire can be lowered if there are no other wires below that. 
With the compression, for spin glass defined on a graph with its maximal degree $D_{\rm max}$ being finite, the total number of physical qubits in this 3D local quantum annealing  architecture is proportional to $N^2$, having the same scaling form as in Lechner-Hauke-Zoller model~\cite{Lechnere1500838}. 
For many difficult NP-problems including unit-disk graph problems~\cite{1990_Clark_UDgraphs}, 
 the most difficult instances of 3SAT~\cite{1993_Crawfor_AAAI}, 
 and partition of power-law degree distributed social networks~\cite{Barabasi}, the degree $D_{\rm max}$ in the corresponding Ising formulation is typically a finite number. 
For Ising models on graphs with different connectivity, we would end up in the compression with the encoding architectures having different geometry. 
While this may be difficult to implement with solid state systems~\cite{2019_Oliver_SCQubits,2013_Eriksson_RMP},  dynamical manipulation of  the geometry can be achieved by controlling lasers with atomic quantum systems  (see Sec.~\ref{sec:exp}).

One property we would like to emphasize about the 3D cubic architecture is that it has the ingredient of repetition error-correcting code due to the logical qubit duplication. This gives the quantum annealing architecture protection against readout errors. We assume an independent single-spin-flip error with probability $\epsilon<50\%$ in the readout. Taking a majority voting scheme for decoding, the probability for the decoded logical bit configuration to be erroneous has an exponentially small upper-bound, 
\be 
p_{\rm logical}  \le  N\exp\left(-ND\left[\frac{\lfloor N /2 \rfloor}{N} \| (1-\epsilon)\right]\right), 
\ee 
with $D(a \| b)$ the standard relative entropy $a\ln(a/b)+(1-a)\ln[(1-a)/(1-b)]$. 
This result is obtained from the Chernoff bound of Binomial distribution of spin flips, which is reasonably tight. The 3D cubic quantum annealing architecture is thus  robust against readout errors.

\subsection{Engineering $m$-body interactions through a quantum wire network} 
\label{sec:mbody} 
Since a direct formulation of many computation problems such as 3SAT and prime factorization requires three-body~\cite{2020_Lin_PRA} or higher-order interactions~\cite{2008_Du_Factorization}, here we  generalize the quantum wire connector protocol to a quantum wire network which mediates $m$-body interactions in general. 
The application of this engineering scheme could reach beyond the present scope of quantum annealer construction, and may be adopted to programmable quantum simulations of novel quantum many-body physics.

We consider three qubits $A$, $B$, and $C$, with the corresponding Pauli operators $\hat{\sigma}^z_{A}$, $\hat{\sigma}^z_B$, $\hat{\sigma}^z _C$.
With an ancilla $\hat{\tau}^z_o$ introduced to connect the three qubits through three Ising ferromagnetic quantum wires [see Fig.~\ref{fig:f1}({\bf b})], we can achieve an interaction, 
\be 
H_{Y} =2J_Y  \hat{ \tau}^z_o \left[ \hat{\sigma}^z_C - \Delta \left( \hat{\sigma}^z _A + \hat{\sigma}^z _B\right) + (1+\Delta) \right], 
\label{eq:HY} 
\ee 
with $J_Y$ parametrizing the overall interaction strength, and $\Delta$ a dimensionless parameter describing the energy gap to be discussed below.  
The connections then make a Y-junction with the qubits $A$, $B$, and $C$ placed at the three ends, and the ancilla ($\hat{\tau}^z_o$) at the middle crossing point.    
Assuming $\Delta >2$, the effective interaction mediated by the ancilla is obtained by projecting to the low-energy subspace~\cite{2002_Hammer_Math} having  
$2 \hat{\tau}^z _o=\hat{\sigma}^z _A\hat{\sigma}^z _B +\hat{\sigma}^z _A + \hat{\sigma}^z _B -1$. 
The states violating this condition have a minimal energy gap of $4(\Delta-2)J_Y$.  
The projection produces,  
\bea 
 H_{Y, {\rm eff}} /J_Y &=& \hat{\sigma}^z _A \hat{\sigma}^z _B \hat{\sigma}^z _C  + (1+\Delta) (\hat{\sigma}^z _A + \hat{\sigma}^z_B) - \hat{\sigma}^z_C \nn \\ 
&+ &\hat{\sigma}^z_A \hat{\sigma}^z_C +\hat{\sigma}^z_B \hat{\sigma}^z _C   + (1-\Delta) \hat{\sigma}^z_A\hat{\sigma}^z _B. 
\eea 
This is a non-separable three-body  Ising-type interaction.  
Combining this Y-junction with another three quantum wires connecting $A$ with $B$, $B$ with $C$, and $C$ with $A$, this permits full programmability of all Ising interactions among the three qubits. 

Since a spin flip in the ancilla leaving the low energy subspace has an energy cost of $4(\Delta-2)J_Y$, 
 the error rate of the Y-junction connector is exponentially suppressed as $\propto e^{-4\beta J_Y(\Delta-2) }$ ($\beta$ is the inverse temperature), provided that the temperature is much lower than  the energy gap.   
In presence of quantum fluctuations for example driven by $\hat{\sigma}_{A,B,C}^x$, the Y-junction connector is valid assuming that the energy gap dominates over other energy scales.

In this context we note that a resolution of four-body Ising interactions into two-body interactions with an auxiliary spin has been discussed in Ref.~\onlinecite{2017_Glaetzle_NC}.
More generally, an $m$-body interaction can be reduced to ($m-1$)-body interactions with a price of introducing one ancilla using Y-junction because with Y-junction any three-body Ising interaction can be induced by two-body terms.  This means $m$-body interactions can be  recursively reduced to two-body eventually. 
The quantum wires mediating $m$-body interactions would then form a complex network, whose topology quickly become highly complex as $m$ increases.  
For the quantum annealer construction, the complex network mediating $m$-body interaction can be  further mapped with the qubit duplication scheme to the non-crossing architecture in three-dimensions just as shown in Fig.~\ref{fig:f1}. 
This implies that a locally coupled spin model on a 3D cubic lattice is sufficient to encode arbitrary all-to-all coupled Ising Hamitonians.

\subsection{Requirement on Ising coupling between duplicated qubits} 
In the 3D  cubic quantum annealer (Eq.~\eqref{eq:Code}), it is required to have $\hat{\sigma}^z_{ik}$ with the same $i$-index ferromagnetically polarized in the ground state of $H_{\rm P}$, because they represent the same logical qubit.  
 In order to enforce this ferromagnetic polarization, it is adequate to choose  
 \be 
 J_d <- K_{\rm max} \times D_{\rm max} , 
 \label{eq:Jdcondition} 
 \ee 
 with $D_{\rm max} $ the maximal degree of $G$. 
 With a finite maximal degree, the required coupling strength for $J_d$ does not increase with qubit number. 
Even with all-to-all couplings, we suggest starting from a small number of $r \equiv |J_d| /K_{\rm max} =2$  and ramping it up until no defect is found in the duplicated qubits because the condition in Eq.~\eqref{eq:Jdcondition} is unnecessary for typical instances, for the argument provided below. Only for the rare worst instances, it is required to set $r> D_{\rm max}$, which may  compromise the physical computation speed by a factor of $1/r$ in experimental implementation.

\begin{figure}
\includegraphics[width=1\linewidth]{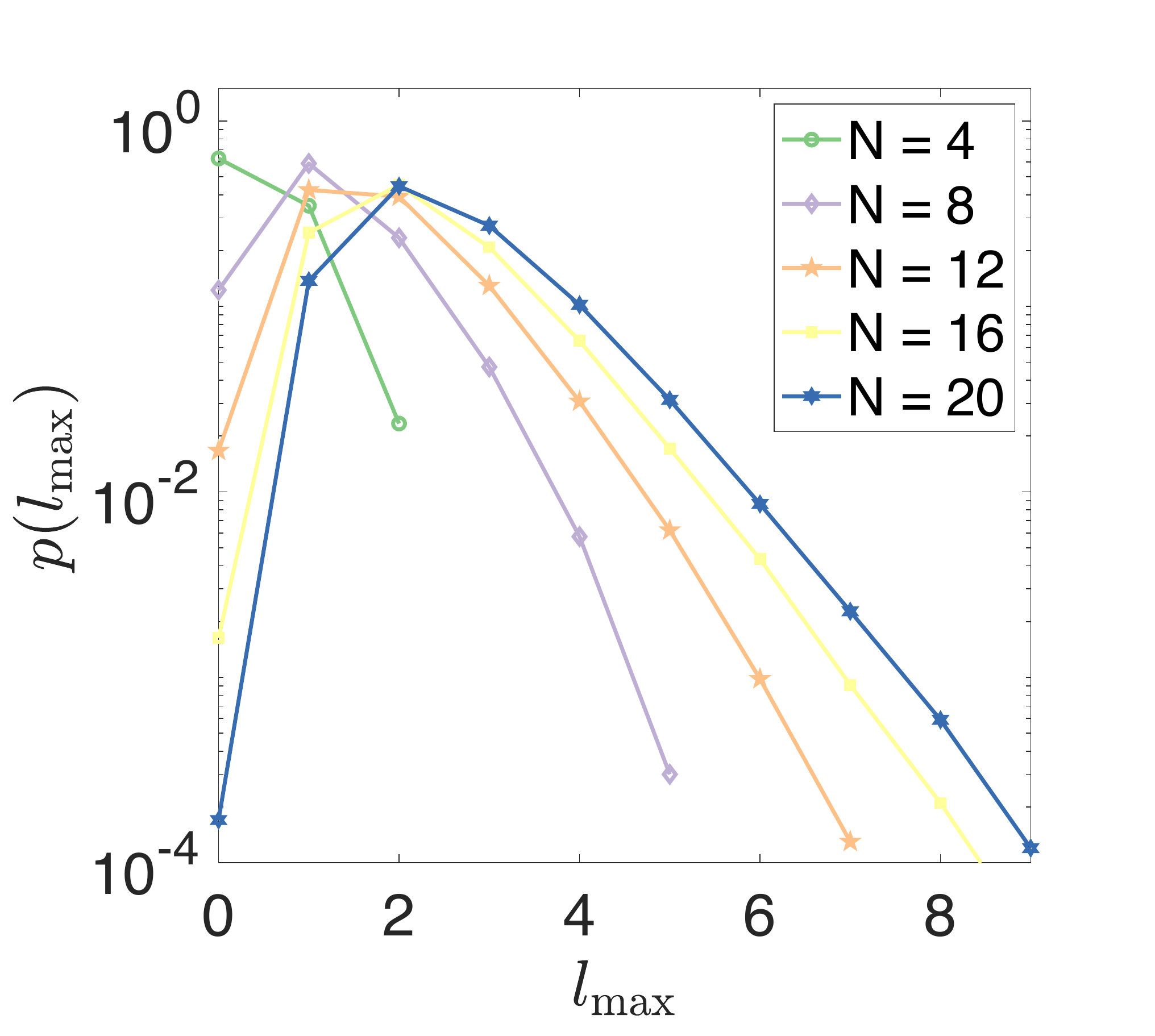}
\caption{
{\bf The probability distribution of $l_{\rm max}$ in Ising spin glass ground states with random couplings.}   
We simulate the Ising spin glass models with all-to-all random couplings with number of vertices $4$, $8$, $12$, $16$, and $20$. The random couplings are drawn from $[-1, 1]$ according to a uniform distribution. We obtain the ground state and calculate the value of $l_{\rm max}$ (see the main text). The statistics is taken over samples of  $10^5$ random problem instances. It is evident that the probability distribution $p(l_{\rm max})$ exhibits an exponential decay at large $l_{\rm max}$. 
}
\label{fig:frustratedist}
\end{figure}

Here, we elaborate on the requirement on $J_d$ to avoid defects in the duplicated qubits. With the condition in Eq.~\eqref{eq:Jdcondition}  satisfied, it is straightforward to show the ground state of the 3D cubic quantum annealer has no defect in the duplicated qubits, because the energy penalty to have a defect induced by $J_d$ is guaranteed to be larger than any possible energy gain. Here we argue that the defects in the duplicated qubits can still be sufficiently suppressed even if this condition is not satisfied. 
Setting $|J_d|/ K_{\rm max}  = r\ge 1$, the energy cost for randomly distributed $N_D$ number of defects is typically $4N_D |J_d|$ and the energy gain is smaller than $2N_D K_{\rm max} $, which means such defects would not  exist in the ground state. More of our concern is about the softer defects of spins flipping in continuous domains. 
Considering spin flips in a continuous domain, $\hat{\sigma}^z_{i_0, k_0}$, $\hat{\sigma}^z_{i_0, k_0+1}$, $\hat{\sigma}^z_{i_0, k_0+2}$, $\ldots$ $\hat{\sigma}^z_{i_0, k_0+l-1}$ with a domain size $l$, the energy cost for such defects is $4|J_d|$ if the domain is not at the boundary and is $2|J_d|$ otherwise.  The maximal energy gain is $2lK_{\rm max}$, which is potentially larger than the energy cost. This would then cause errors in the 3D cubic quantum annealer as those defects are no longer energetically suppressed. However the maximal energy gain is only reached for rare cases with all terms 
of $K_{i_0 i'} s_{i_0} s_{i'} $ [$(i_0+i') \, {\rm mod}\, N = k_0, k_0+1, \ldots k_0+l-1$] being coherently positive in the actual ground state of the Hamiltonian in Eq.~\eqref{eq:IsingHam}. Note that the link $(ii')$ with $K_{i i'} s_i s_{i'} >0$ in the ground state are frustrated. Such links do not exist in frustration-free models. In general without fine tuning, the probability for that maximal energy gain to really happen should decay exponentially with $l$.  With the choice of $|J_d|/ K_{\rm max}  = r$, the probability for the 3D cubic quantum annealer to be erroneous is thus expected to decay exponentially with $r$. This argument is further confirmed with numerical simulations of Ising spin glass with all-to-all random couplings (Fig.~\ref{fig:frustratedist}), where the couplings $K_{ii'}$ are drawn from $[-1, 1]$ according to a uniform distribution. Given a ground state configuration, for each logical qubit index $i_0$, we define a maximal domain size $l_{\rm max}$ to be a maximal value of $l$ that has all  the links $(i_0i')$ with $[(i_0+i') \, {\rm mod}\, N = k_0, k_0+1, \ldots k_0+l-1$] being frustrated ($i_0$ and $k_0$ can be arbitrarily chosen).  
 In Fig.~\ref{fig:frustratedist}, we show the probability distribution of $l_{\rm max}$ obtained from numerical simulations, and  confirm that the probability distribution exhibits an exponential decay.  The proposed 3D cubic architecture should thus provide a scalable quantum annealing Hamiltonian.

\medskip 
 \section{Finite temperature effect} 
 \label{sec:finitetemp}

Since  the physical devices performing quantum annealing may not operate at absolute zero temperature, we describe finite temperature effects in this section. 
In the following, we show that the scheme of quantum wire mediating long-range interaction provided in Sec.~\ref{sec:HamConstruction}  has reasonable scalability even taking into account thermal excitations. This is particularly crucial to optical lattice experimental implementation to be discussed in Sec.~\ref{sec:exp}. 
Moreover, the scheme can be adapted to construction of an annealer producing finite temperature ensemble.

\begin{figure}
\vspace{0.5cm} 
\includegraphics[width=.85\linewidth]{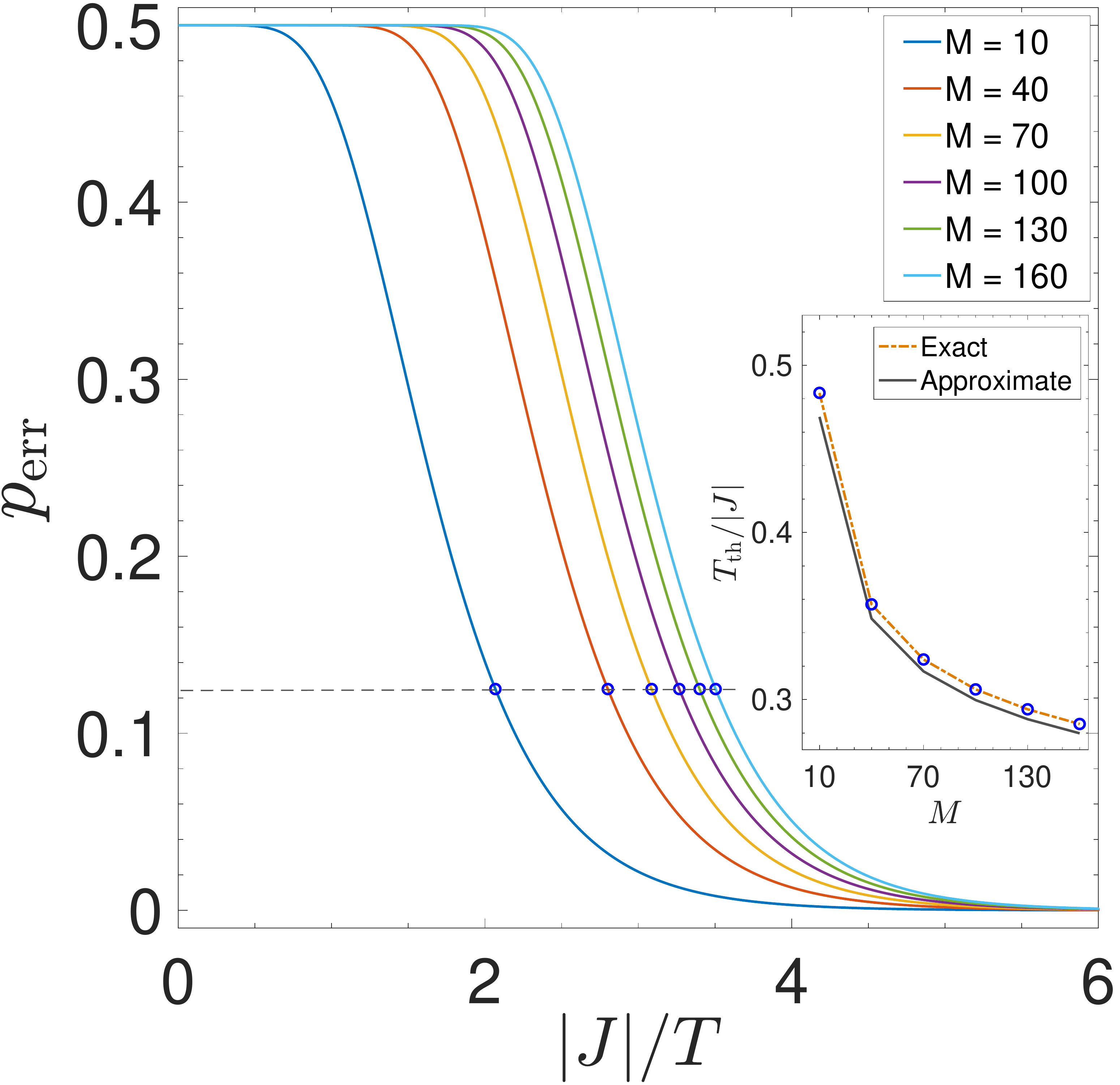}
\caption{
{\bf The error rate ($p_{\rm err}$) in the ferromagnetic quantum wire at finite temperature ($T$).}   
We choose a series of wire lengths, $M = 10, 40, 70, 100, 130, 160$. 
The gray dash line indicate a threshold probability  of $p_{\rm th}=1/8$. 
The inset shows the corresponding temperature threshold $T_{\rm th}$ with different $M$, calculated from  the exact expression in Eq.~\eqref{eq:Tth_exact} and the approximation in Eq.~\eqref{eq:Tth_approximate}.  
}
\label{fig:defect}
\end{figure}

 \subsection{Thermal defect causing error in the ferromagnetic quantum wire} 
Considering finite temperature effect, the thermally excited domain wall defects in the ancilla quantum wires may cause errors. We emphasize here that the error probability corresponds to having odd number of domain walls in the Ising ferromagnetic quantum wire, because the effective coupling through a ferromagnetic wire  with even number of domain wall defects is equivalent to  a wire in its ground state having no defect. 
In presence of an even number of domain wall defects, the two spins at the ends are still ferromagnetically oriented as in the ground state, whereas their orientation is antiferromagnetic  for an odd number of domain wall defects. 
In general, the ferromagnetic quantum wire with $M$ ancilla that couples two distant qubits, is described by the Hamiltonian $\hat{H}_{J} =J\sum^{M-2}_{j=0}\hat{\tau}^z_j\hat{\tau}^z_{j+1}$.  The thermal ensemble of these ancilla is given by the density matrix operator, $\hat{\rho} = \exp(-\beta \hat{H}_{J} )/{\rm Tr}[\exp(-\beta \hat{H}_{J})] $, with $\beta$ the inverse temperature. The Boltzmann constant is set as a unit throughout. 
 
The probability of having odd number of domain wall defects ($p_{\rm err}$) corresponds to the probability of the two end ancilla of the quantum wire being opposite. 
With a transfer matrix method~\cite{Altland-Simons},  
we obtain the error rate as 
\be
p_\text{err}=\frac{[\lambda_+(J)]^{M-1} - [\lambda_-(J)]^{M-1}}{2[\lambda_+(J)]^{M-1}},
\label{eq:perr} 
\ee
here $\lambda_\pm(J)=e^{-\beta J}\pm e^{\beta J}$. 

At a request of error rate below a certain threshold $p_{\rm th}$, we let $p_{\rm{err}} < p_{\rm th}$, which leads to a requirement on the temperature 
$T<T_{\rm th}$, with  the temperature threshold, 
\be
T_{\rm th} =2 |J|  \left[ \ln\frac{1+(1-2p_{\rm th})^{\frac{1}{M-1}}}{1-(1-2p_{\rm th})^{\frac{1}{M-1}}} \right]^{-1}.
\label{eq:Tth_exact}
\ee 
Considering a small error rate threshold, we have $(1-2p_{\rm th})^{\frac{1}{M-1}}\approx 1-\frac{2p_{\rm th}}{M-1}$, and the temperature threshold takes a more illuminating form of,  
\bea
\label{eq:Tth_approximate}
T_{\rm th}\approx 2|J| \left[\ln\frac{M-1-p_{\rm th}}{p_{\rm th}}\right]^{-1}. 
\eea
Therefore, the temperature threshold $T_{\rm th}$ decreases logarithmically with the length of the quantum wire $M$, 
or the distance of spins we are aiming to couple. 
We conclude that the proposed 3D cubic quantum annealing architecture is reasonably scalable even taking into account the potential finite temperature effect in the experimental implementation. 

For a number of choices of quantum wire lengths from $10$ to $160$, the numerical values of error rate $p_{\rm err}$ and the temperature threshold $T_{\rm th}$ setting $p_{\rm th}= 1/8$ are shown in Fig.~\ref{fig:defect}. The temperature threshold is found to be at the order of several tenths of $J$.

\begin{figure*}
\begin{center}
\includegraphics[width=\linewidth]{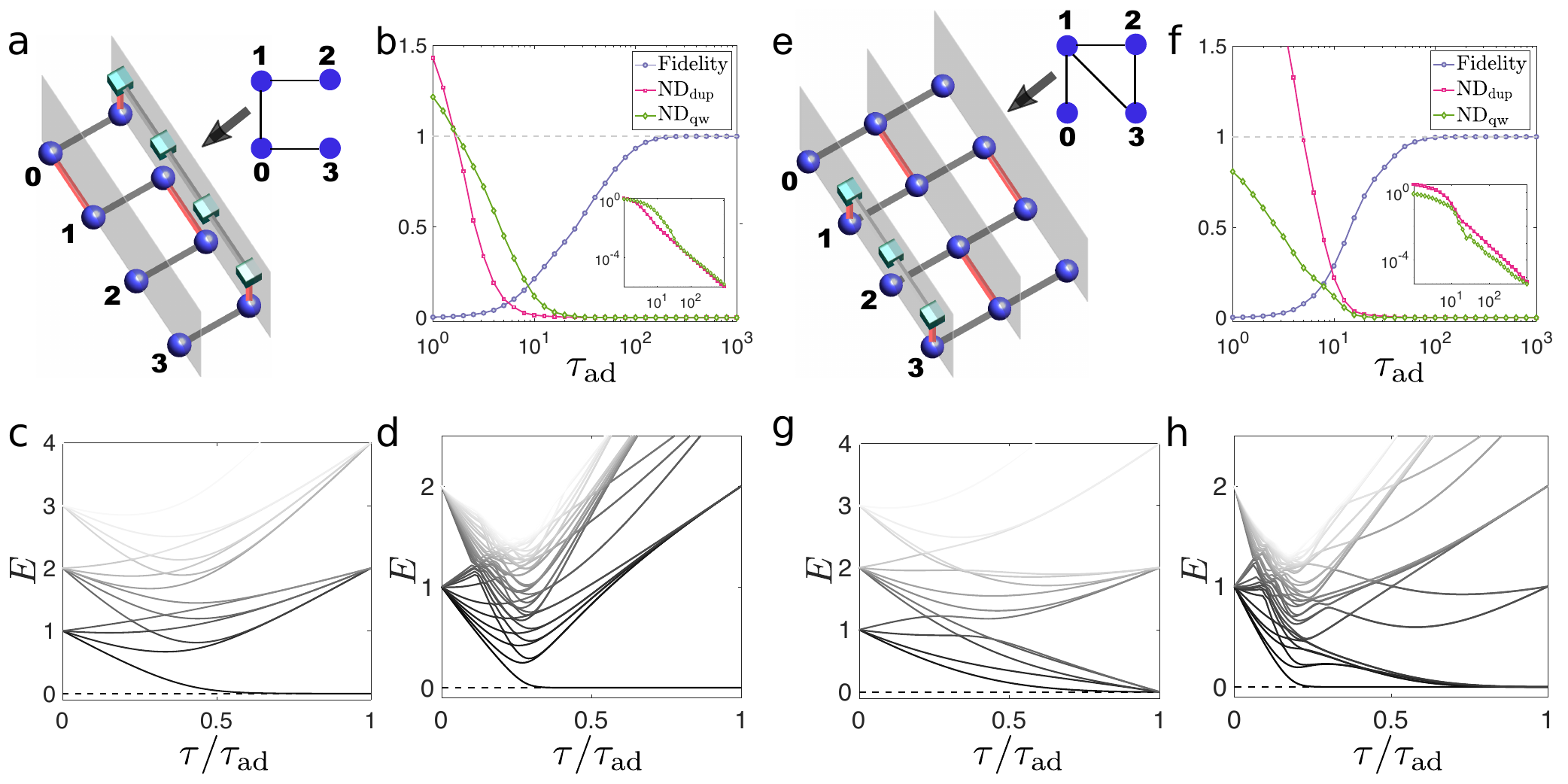}
\end{center}
\caption{{\bf Demonstration of the 3D cubic architecture for quantum annealing applied to Max-Cut problems.} 
{
For the graph shown at the top right of ({\bf a}, {\bf e}), we encode the Max-Cut problem of the graph in a spin glass model. The four logical qubits correspond to the four vertices in the graph. The corresponding 3D encoding architecture is shown in ({\bf a}, {\bf e}), with the unnecessary layers removed. Its numerical performance is shown in  ({\bf b}) and ({\bf f}), respectively, taking the local quantum annealing Hamiltonian in Eq.~\eqref{eq:Code}. 
The dependence of fidelity, and the averaged number of defects in the duplication qubits ${\rm ND}_{\rm dup}$ and in the connecting quantum wires ${\rm ND}_{\rm qw}$, on the total adiabatic time $\tau_{\rm ad} $ are shown ({see main text}). 
The insets show the monotonic decay behavior of ND$_{\rm dup}$ and ND$_{\rm qw}$ at large $\tau_{\rm ad}$. For both cases, the quantum annealer reaches a fidelity of $50\%$ around $\tau_{\rm ad} =20$. The defect numbers ${\rm ND}_{\rm dup}$ and ${\rm ND}_{\rm qw}$  drop down to below one percent at about $\tau_{\rm ad} = 30$, and decreases rapidly beyond that. 
The corresponding instantaneous eigenstate energy spectra for the lowest forty states are shown in ({\bf d}) and ({\bf h}). 
For comparison, the energy spectra with the direct non-local Hamiltonian in Eq.~\eqref{eq:qIsingHam} are shown in ({\bf c}) and   ({\bf g}). 
The energy spectra are all shifted with respect to the instantaneous ground state energy, and the dashed lines thus represent the ground state levels in ({\bf c}, {\bf d}, {\bf g}, {\bf h}). 
The coupling in the Ising formulation of the Max-Cut problems is taken as an energy unit (see main text).  
We choose the parameters $J_d = -1.1D_{\rm max}$ and $J = -1.5$. 
}
}
\label{fig:f2}
\end{figure*}

\subsection{Construction of a finite temperature annealer} 
We further point out that  the idea of using ferromagnetic quantum wires to couple distant logical qubits also applies to constructing an annealer that produces finite temperature ensemble. 
For each pair of spins $(ii')$, we have a ferromagnetic quantum wire with $M_{ii'}$ ancilla that couples to $\hat{\sigma}^z_i$ and $\hat{\sigma}^z_{i'}$, described by the Hamiltonian in Eq.~\eqref{eq:IsingSpinChain}. 
The thermal fluctuations are described by  a density matrix operator  
$ 
\hat{\rho}_{\rm QWC} \propto \exp \left( -\beta \hat{H}_{{\rm QWC}, ii'}  \right). 
$
The induced couplings between $\hat{\sigma}^z_i$ and $\hat{\sigma}^z_{i'}$ are derived by tracing out the fluctuations of the ancilla,
$
\hat{\rho}_{\rm eff } = {\rm Tr}_{\rm ancilla}  \left [\hat{ \rho}_{\rm QWC}\right] .  
$ 
To produce a thermal ensemble distribution determined by the Hamiltonian in Eq.~\eqref{eq:IsingHam}, 
we require 
$
\hat{ \rho} _{\rm eff}  = Z_0 \exp\left(  -\beta K_{ii'} \hat{\sigma}^z _i \hat{\sigma}^z_{i'} \right), 
$
with $Z_0$ some arbitrary constant. 
With the transfer matrix method~\cite{Altland-Simons} we obtain, 
\begin{widetext} 
\be 
\exp\left( 2\beta K \right) = 
\frac{\lambda_+ (J_L ) \lambda_+ (J_R) [\lambda_+ (J) ]^{M-1}  - 
	\lambda_- (J_L ) \lambda_- (J_R)  [\lambda_-  (J)]  ^{M-1}  }
	 {\lambda_+ (J_L ) \lambda_+ (J_R) [\lambda_+  (J)]^{M-1} + 
	\lambda_- (J_L ) \lambda_- (J_R) [\lambda_-  (J)]^{M-1} } . 
\label{eq:KJrelation} 
\ee 
\end{widetext} 
Here, the subscripts `$_{ii'}$' labeling different quantum wires  are suppressed in $K$, $J_L$, $J_R$ and $M$, to save writing. 
At zero temperature limit,  
this result agrees with  the domain wall construction that considers the ground state directly in Sec.~\ref{sec:HamConstruction}.

At finite temperature, Eq.~\eqref{eq:KJrelation} is satisfied by taking 
\bea 
&& J_{ii'}  =-|K_{ii'} | - \ln \left[ 2(M_{ii'} -1)\right]/2\beta, \nn \\  
&& J_{L,ii'}  /K_{ii'} = -J_{R,ii'} /|K_{ii'}| = -\frac{1}{2\beta |K_{ii'}|} \ln \frac{1-C_{ii'}}{1+C_{ii'}}, 
\eea 
with 
$
\textstyle C_{ii'} = \left(\frac{\lambda_- (-|{{ {K_{ii'}}}}|) }{\lambda_+(-|{{ {K_{ii'}}}}|)} 
\left[\frac{\lambda_+(J_{ii'})  }{ \lambda_-(J_{ii'})}\right]^{M_{ii'}-1}\right)^\frac{1}{2}. \nn 
$
This means the required coupling strength in the quantum wire increases logarithmically with the length $M_{ii'}$, 
and that the magnitudes of $J_L$ and $J_R$ decrease monotonically with $M_{ii'}$ and $\beta$, having a lower bound, $|K_{ii'}|$.
We thus conclude that the 3D cubic architecture is adaptable for finite temperature annealing.

\medskip 
 \section{Numerical  demonstration} 
 \label{sec:demo} 

 For demonstration the proposed quantum annealing architecture is now applied to Max-Cut and prime factorization problems. 
 In our protocol, it is required to have no defects in the duplicated logical qubits and in the quantum wire connectors. 
 We thus simulate the quantum annealing process and check   
  the number of defects in the duplicated qubits, 
\be 
{\rm ND}_{\rm dup} =\frac{1}{2} \sum_i \sum_k \left[1-\langle \hat{\sigma}^z_{ik} \hat{\sigma}^z_{i,k+1} \rangle \right], 
\ee 
and the number of defects in the quantum wires, 
\be 
{\rm ND}_{\rm qw}  = \frac{1}{2} \sum_{(ii')\in E} \sum_j
			 \left[		1-\langle \hat{\tau}^z_{(ii'),j} \hat{\tau}^z_{ (ii'), j+1} \rangle \right]. 
\ee
In our simulation of the quantum dynamics, we set the Planck constant $\hbar$ as a unit.  
We confirm that these defects are indeed suppressed in the quantum adiabatic evolution solving  both Max-Cut and prime factorization problems. 
The simulation is carried out for very small system sizes for demonstration. The investigated problems could also be examples of demonstration experiments for small scale quantum hardware.

\subsection{Solving Max-Cut problems with the 3D cubic architecture} 
As a concrete demonstration, we simulate the performance of the  3D cubic  quantum annealing architecture  in solving Max-Cut problems. Given a graph $G$ and an integer $P$, the Max-Cut problem is to determine whether there is bipartition of the graph that breaks at least $P$ edges. This problem is NP complete. Finding the partition breaking the most edges is NP hard~\cite{2014_Lucas_FIP}.  This problem arises in a broad range of applications including financial portfolio optimization and social network analysis~\cite{2014_Lucas_FIP,Barabasi}. The fact that solving Max-Cut on a classical computer is exponentially hard but its verification is easy, makes it an ideal arena for quantum annealing to demonstrate applicational quantum advantage.  

The spin glass Hamiltonian encoding the Max-Cut problem is the one in Eq.~\eqref{eq:IsingHam}, with $K_{ii'} $ replaced with a constant positive value $K$ to be set as $1$ 
{(the energy unit for the numerical results in Fig.~\ref{fig:f2})}, and the local fields $b_i=0$. 
Following the standard quantum adiabatic computing, the initial state of the quantum annealer is set to be the ground state of 
an initial Hamiltonian $\hat{H}_0$ (Eq.~\eqref{eq:Code}). 
The two example graphs are studied in Fig.~\ref{fig:f2}. 
The  expectation value  of the number of defects in the duplicated qubits and in the quantum wire connectors with respect to the final quantum states, ${\rm ND}_{\rm dup}$ and ${\rm ND}_{\rm qw}$, are shown in Fig.~\ref{fig:f2}. 
The undesired defects are indeed greatly suppressed at a reasonable adiabatic evolution time. At long time, the time dependence of the defect numbers approach to a monotonic power law decay as expected from Kibble-Zurek theory~\cite{Kibble_1976}. For both studied graphs with [Fig.~\ref{fig:f2} ({\bf c}, {\bf d})] and {without [Fig.~\ref{fig:f2} ({\bf a}, {\bf b})] loops}, the quantum annealer reaches the correct solution of the Max-Cut problem with finite fidelity at  a reasonable time cost.  
{
For comparison we also calculate the instantaneous energy spectra for  the direct nonlocal quantum annealing model in Eq.~\eqref{eq:qIsingHam}  and for the local quantum annealing in Eq.~\eqref{eq:Code}. 
For the studied Max-Cut problems, the energy gap above the ground state manifold of the latter is comparable to the former. 
Considering more difficult  problems on larger graphs, eliminating the domain wall defects  in the 3D local quantum annealer in the adiabatic quantum evolution would cause an  increase in the time cost overhead, which we expect to scale polynomially with qubit number for the Kibble Zurek scaling~\cite{Kibble_1976}. 
}

\begin{figure}
\begin{center}
\includegraphics[width=\linewidth]{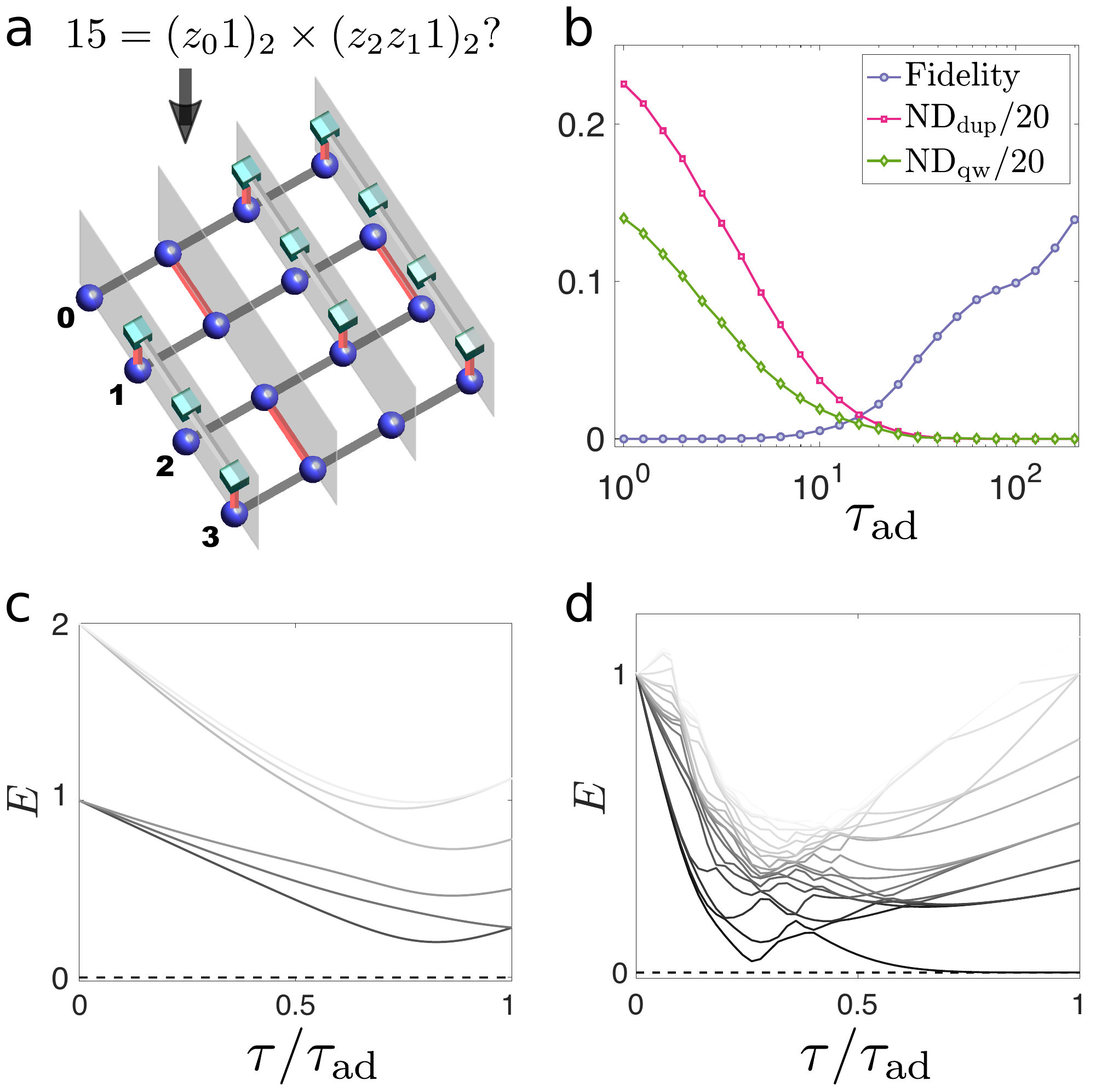}
\end{center}
\caption{ 
{\bf Demonstration of the 3D cubic architecture for quantum annealing applied to prime factorization.}  
{
A prime factorization of $15$ is investigated. This problem is mapped to a spin glass model of four vertices with all-to-all couplings.  ({\bf a}) shows the 3D cubic quantum annealing architecture having $26$ qubits. ({\bf b}), the dependence of fidelity, and the averaged number of defects in the duplication qubits ${\rm ND}_{\rm dup}$ and in the quantum wire connectors ${\rm ND}_{\rm qw}$, on the total adiabatic time $\tau_{\rm ad} $.
The defect numbers  ${\rm ND}_{\rm dup}$ and ${\rm ND}_{\rm qw}$ are rescaled for cosmetic reasons. 
Here we choose the parameters $J_d = -1.1D_{\rm max}K_{\rm max} $ and $J = -1.5 K_{\rm max}$. 
The defect excitations become negligible for an adiabatic time $\tau_{\rm ad}>20$. 
({\bf d}), the instantaneous eigenstate energy spectra of the local quantum annealing Hamiltonian in Eq.~\eqref{eq:Code}. 
We calculate the lowest twenty states. 
({\bf c}), the energy spectra for quantum annealing with the direct nonlocal three-body Hamiltonian (see Eq.~\eqref{eq:Hfactordirect}). 
The energy spectra are shifted with respect to the instantaneous ground state energy, and the dashed lines thus represent the ground state levels in ({\bf c}, {\bf d}).  
The coupling strength of the three body interaction is taken as an energy unit. 
} 
}
\label{fig:f3}
\end{figure}

\subsection{Factorization with the 3D cubic architecture} 

We further demonstrate that the 3D cubic quantum annealing architecture can be used to solve prime factorization which has important implications in cryptography for its lack of efficient algorithms with classical computing. It has been shown that the problem of factorization can be solved by adiabatic quantum computing~\cite{2008_Du_Factorization}. In our demonstration, we examine the factorization of $15 = p\times q$. With the binary representation $p = (z_0 1)_2$, $q = (z_2 z_1 1)_2$, 
finding the solution of $(p, q)$ is equivalent to an optimization problem minimizing $(15-p\times q)^2$. 
Through the binary representation, this optimization problem is encoded as solving the ground state of the Hamiltonian,  
\bea 
\label{eq:Hfactordirect} 
 H_{p} /\varepsilon &=&  32 z_0 z_1 z_2 -14z_0 z_2 -12 z_0z_1+ 4z_1 z_2 \nn \\ 
& -& 13 z_0 - 13 z_2 -24 z_1,    
\eea 
with $\varepsilon$ a rescaling factor, set as $1/32$ in our numerical simulation.  
The coupling strength of the three body term $z_0z_1z_2$ then sets an energy unit.
The $z$ variables take boolean values, $0$ and $1$, and are related to Ising spins by $z_i = (s_i+1)/2$.
The three-body term  is then reduced to a quadratic form using the Y-junction approach in Sec.~\ref{sec:mbody},
where we set  the dimensionless parameter $\Delta$ in Eq.~\eqref{eq:HY} as  $\Delta =4$. 
The resultant  quadratic Ising spin glass model is further mapped to the local quantum annealing architecture (Eq.~\eqref{eq:Code}),  with the scheme developed in Sec.~\ref{sec:HamConstruction}. 
The corresponding 3D encoding architecture has $26$ physical qubits {[Fig.~\ref{fig:f3} ({\bf a})]} and is not compressible.

{Fig.~\ref{fig:f3} ({\bf b})} shows the performance of the 3D local quantum annealing for the prime factorization. With the adiabatic time about $\tau_{\rm ad} >20$, the defect excitations in the quantum wire connectors or the duplicated qubits become negligible. In the whole region of $\tau_{\rm ad} $ we have investigated, we find monotonic increase of the fidelity, which reaches $10\%$ at about $\tau_{\rm ad} = 100$. 
The instantaneous energy spectra for  the direct quantum annealing of the non-local three-body model in Eq.~\eqref{eq:Hfactordirect}  and for the local quantum annealing in Eq.~\eqref{eq:Code} are shown in Fig.~\ref{fig:f3} ({\bf c}, {\bf d}), respectively. 
The local quantum annealer has an energy gap above the ground state about one sixth of the non-local three-body model. This local encoding thus introduces an overhead in the time cost considering the physics  of Landau Zener transition~\cite{2008_Chakrabarti_RMP,2018_Lidar_RMP}. In our numerics, we find that the limitation in the performance of the 3D local quantum annealing  is mainly from domain wall defects [Fig.~\ref{fig:f3}({\bf b})], 
whose number  should follow a polynomial Kibble-Zurek scaling~\cite{Kibble_1976,1985_Zurek}. 
We thus expect the overhead scales polynomially with qubit number, 
which is acceptable because an exponential quantum speedup with the  quantum annealer is expected~\cite{2008_Du_Factorization,Kais_2018}.

{\it Remark.---} 
In the above demonstration including Max-Cut and prime factorization, we choose a linear schedule in the adiabatic quantum computing (Eq.~\eqref{eq:Code}), which may not have the best performance in terms of computation time. This time cost can be dramatically improved by optimizing the schedule~\cite{2020_Lin_PRA,2020_Hsieh_arXiv}, or adding counterdiabatic drivings~\cite{2003_Demirplak_JPCA,2009_Berry_JPA,2017_Polkovnikov_PR}.

\medskip 
\section{Experimental candidates} 
\label{sec:exp}

For an experimental realization of the proposed quantum annealing architecture (see Eq.~\eqref{eq:Code}), 
we focus on atomic systems for which 3D arrays including Ising quantum wires can be arranged with laser-created optical lattices~\cite{gross2017quantum} or tweezers~\cite{browaeys2020many}.
In these systems, the dynamical manipulation of the geometry of qubits can be achieved by controlling lasers. 
Below we outline a physical implementation of the quantum annealer with superexchange in atomic Hubbard models and Rydberg $p$-wave dressing interactions for ultracold atoms confined in optical lattices.

\subsection{Interaction design through atomic  superexchange in an optical lattice} 
We consider atoms confined in a 3D optical lattice forming a Mott insulating state, with the atomic internal states encoding the qubits. 
With a far-detuned optical lattice, the 
familiar  atomic superexchange is Heisenberg interaction~\cite{2003_Duan_PRL}. 
One approach to introduce Ising spin interactions as required  in the quantum annealing model (see Eq.~\eqref{eq:Code}) is to freeze the tunneling of one spin component, say spin $\ket{\down}$, with a spin-dependent lattice  potential, 
$
 V_{\sigma} (x) = {(V_0+\sigma V_1)} \sum_{\nu=1,2,3} \sin^2 (\pi x_\nu), 
$ 
where $\sigma = \pm $ representing two spin (or hyperfine) states of atoms confined in the lattice, and $x_{1,2,3} $ three spatial coordinates~\cite{1999_Jaksch_PRL,2004_Vincent_PRA,2007_Porto_PRL,2008_Zoller_PRL,2017_Pan_ToricCode}. 
 The quantum dynamics is then characterized by the single particle tunneling of spin $\ket{\up}$ atoms, 
 $t_{\up}$, 
the intra-species Hubbard interaction, $U_{\up \up}$ ($U_{\down\down}$), between spin up (down), and the inter-species interaction, $U_{\up \down}$. 
The  virtual tunneling induced superexchange  for bosonic atoms has a coupling strength, 
$J_{\rm ex}  ={  \frac{t_{\up}^2}{2 U_{\up \down}} - \frac{t_\up ^2}{U_{\up\up} } }$~\cite{2003_Duan_PRL}, 
which is ferromagnetic  with $U_{\up \down} >U_{\up\up}/2$, and antiferromagnetic otherwise. 
 The interaction in the lattice are related to the s-wave scattering lengths $a_{\sigma\sigma'} $ by 
$ 
\textstyle U_{\sigma\sigma'} = 4\pi \frac{a_{\sigma\sigma'} \hbar^2} {M} \int d^3 x |w_\sigma(x)|^2 |w_{\sigma'} (x)|^2, 
$  
with $w_\sigma(x)$ the Wannier function for the $\sigma$-component, which depends on the local optical potential. 
Using Feshbach resonances~\cite{2010_Chin_RMP}, 
we tune the external magnetic field to adjust the scattering lengths such that ${a_{\up\down} = a_{\up\up}/2}$, 
and control the superexchange by manipulating the Wannier functions through adjusting  local optical potential with digital micro-mirror devices or related techniques~\cite{gauthier2016direct,2016_Weiss_Science,mazurenko2017cold}. 
The  form of the demanded optical potential can be calculated using our recently developed algorithms which are highly efficient on a classical computer~\cite{qiu2020precise}. 
In this way, both ferromagnetic and antiferromagnetic couplings can be locally achieved.

For alkali atoms a spin dependent lattice for ground state atoms can be created by coupling to  the D line $P$-states with circularly polarized light~\cite{1999_Jaksch_PRL,2004_Vincent_PRA}. This comes with a requirement that the fine structure splitting between D1 and D2 lines should be sufficiently large in order to have both strong enough spin dependence in the optical potential and a sufficiently suppressed spontaneous emission rate. 
We thus consider $^{39}$K atoms whose D1-D2 splitting is about  $2\pi \times 2 $THz. 
{
For K atoms,  
the superexchange at a lattice depth of ten times of recoil energy is estimated to be $2\pi \times 30$ Hz assuming the Hubbard interaction is ten times of the single-particle tunneling, and the corresponding lattice induced spontaneous emission rate is below one tenth Hz.} 
A temperature requirement is set by considering the thermally activated errors in the quantum wire connectors. 
For example, an error rate below $1\%$ requires the temperature below $2\pi\times 10$ Hz for a quantum wire with length $M = 100$, as obtained from Eq.~\eqref{eq:perr}.  
We caution here that the potential challenge with alkali atoms could arise from the spin-dependent lattice induced heating, which should be investigated in experimental studies.

For alkaline earth atoms the spin can be encoded in long lived clock states~\cite{2008_Zoller_PRL,2019_Rey_PRL}. A spin dependent optical lattice realizing the Ising interaction can then be implemented due to the different AC polarizabilties of the ground and excited atomic states. 
In this system, the Hubbard interaction is controllable with an orbital Feshbach resonance~\cite{2015_Zhang_PRL,2015_Folling_PRL}.  
With fermionic $^{87}$Sr atoms, an Ising superexchange interaction can be made to the order $2\pi \times 10$Hz~\cite{2019_Rey_PRL}, which can be tuned to be either ferromagnetic or antiferromagnetic using the spin-orbit coupling techniques achieved in present experiments~\cite{2017_Ye_Kolkowitz_Nature}. 
The experimental system has a quantum coherence time of about $10$ seconds~\cite{2018_Ye_Nature}, for which the computation problems studied in Sec.~\ref{sec:demo} can be experimentally tested.  For larger-size computation problems, it is helpful to consider  performing optimization over the Hamiltonian evolution path~\cite{2020_Lin_PRA,2020_Hsieh_arXiv,2003_Demirplak_JPCA,2009_Berry_JPA,2017_Polkovnikov_PR} or adopting the  iterative quantum annealing approach~\cite{2019_Grass_PRL}.

Our 3D cubic encoding protocol implies that a cubic optical lattice with a spatially-controllable potential would support programmable quantum annealing, although a 3D holographic control over the optical potential still requires further technological developments.

\begin{figure}
\includegraphics[width=\linewidth]{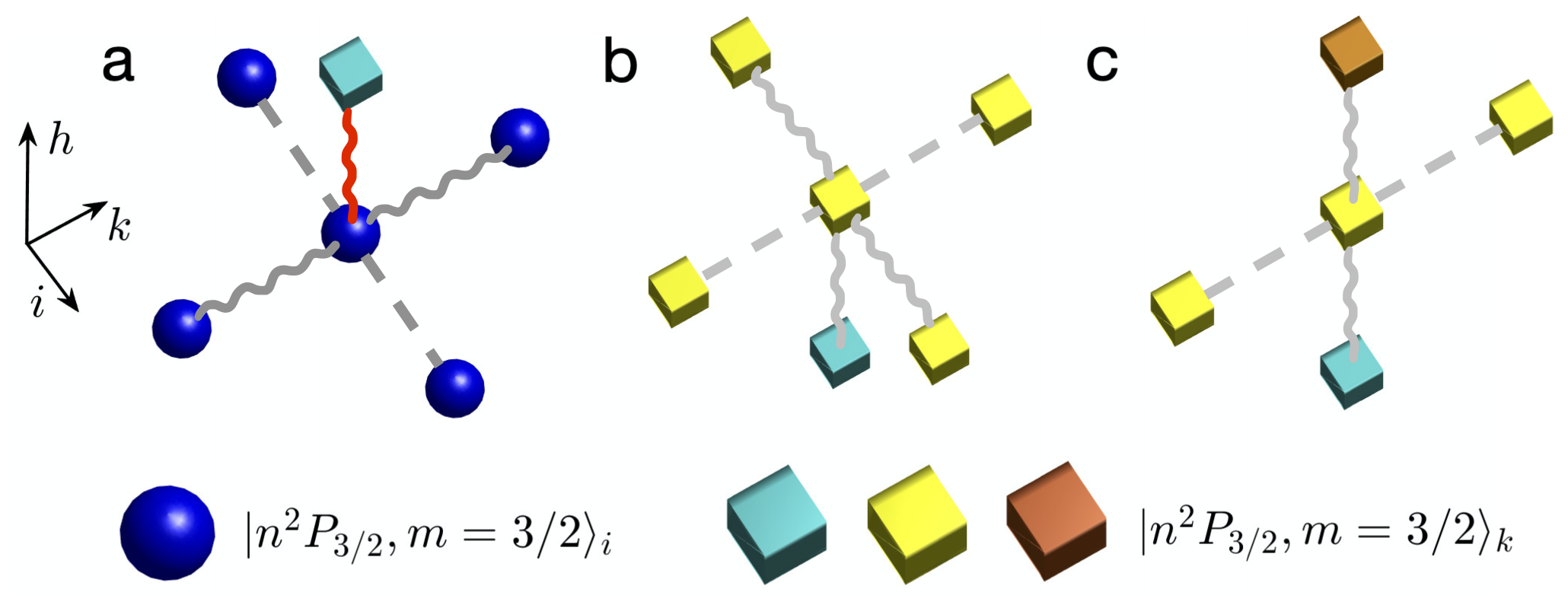}
\caption{
{\bf Illustration of Rydberg $p$-wave building blocks for implementation of the 3D cubic quantum annealing architecture.}
The atoms at the sites of spheres and cubes are dressed by the Rydberg states $\ket{n{}^2P_{3/2},m=3/2}_i$ and $\ket{n{}^2P_{3/2},m=3/2}_k$, respectively, with the subscripts $i$ and $k$ denoting the quantization axes. 
Solid (dashed) lines indicate the presence (absence) of  couplings between the two  Rydberg dressed atoms.
For the angular dependence of Rydberg $p$-wave interaction (Eq.~\eqref{eq:RydbergInt}), 
in the horizontal $ik$-plane the duplicated logical qubits (ancilla) only couple to each other along the $k$ ($i$) direction as shown in ({\bf a}, {\bf b}). Along the vertical $h$-direction, both the duplicated logical qubits and the ancilla interact with their neighbors, as shown in ({\bf a,~b,~c}). 
The red link representing a programmable Ising coupling is realized by the mixed interaction between the two different Rydberg states. 
This coupling orientation corresponds to $(\theta=\pi/2,\phi=\pi/2)$ in the angular dependence of the mixed interaction. 
}
\label{fig:Rydberg}
\end{figure}

\subsection{Rydberg dressing scheme}

 Ising spin interactions  in the quantum annealing architecture can also be implemented with laser excited Rydberg atoms stored in tweezer arrays~\cite{browaeys2020many} or optical lattices~\cite{gross2017quantum}. In addition, a proper choice of the Rydberg state allows an anisotropic, directional Ising interactions which allows to minimize the cross talk between the quantum wires. 
In the quantum annealing architecture [Fig.~\ref{fig:f1} ({\bf c})],  the required Ising interactions are highly anisotropic---quantum wires in different $k$-layers must be decoupled, and most of the duplicated  logical qubits are decoupled along the $i$-direction. To fulfill the anisotropy, it is natural to consider Rydberg $p$-wave interaction~\cite{glaetzle2014quantum}. 

Specifically, we consider the hyperfine states $\ket{ 5{}^2S_{1/2}, F= 2, m_F = 0 }$ and $\ket{ 5{}^2S_{1/2}, F=1, m_F = 0 } $  of $^{87}\text{Rb}$ atoms for qubit encoding as spin $\ket{\uparrow}$ and $\ket{\downarrow}$, respectively.  
We consider a Rydberg dressing scheme~\cite{2010_Buchler_PRL,2010_Pupillo_PRL} with the $\ket{\uparrow}$ state selectively dressed with a Rydberg $p$-state via a circularly polarized laser light to introduce interactions between nearest neighboring atoms in $\ket{\uparrow}$, realizing the Ising couplings in Eq.~\eqref{eq:Code}.  The local fields are controllable by adjusting the laser detuning.  
In this scheme, we use two $p$-states $\ket{r_\medbullet}=\ket{n{}^2P_{3/2},m=3/2}_i$ and $\ket{r_\blacksquare}=\ket{n{}^2P_{3/2},m=3/2}_k$,   $i$ and $k$ indicating the quantization axes (see Fig.~\ref{fig:Rydberg})~\cite{glaetzle2014quantum}. 
Since the interaction programmability requires local controllability of Rydberg dressing, it is experimentally more convenient to confine cold atoms in a lattice with a large lattice constant, for example  about a few microns as used to perform single qubit gates on individual atoms~\cite{2016_Weiss_Science}. 
The angle dependence of van der Waals interactions between atoms dressed with $\ket{r_\medbullet}$ and with $\ket{r_\blacksquare}$,  $V_{\medbullet\medbullet}$ and $V_{\blacksquare\blacksquare}$ respectively, is given in Ref.~\cite{glaetzle2014quantum}. The key feature in these interactions as is relevant to our quantum annealer implementation here is their anisotropy---$V_{\medbullet\medbullet}$ and $V_{\blacksquare\blacksquare}$ vanish along the $i$, and $k$ directions, respectively. The mixed interaction $V_{\medbullet\blacksquare}$, between the two Rydberg $p$ states takes the form of 
\bea
\label{eq:RydbergInt} 
V_{\medbullet\blacksquare} &\sim& \frac{n^{11}}{r^6}A(\theta,\phi). 
\eea
Here, the spherical coordinates $(r,\theta,\phi)$ denote the relative position between two atoms, where the north pole corresponds to the  $+i$ direction, the azimuthal angle measures the direction in the $kh$-plane (Fig.~\ref{fig:Rydberg}). 
The angular part of the mixed interaction $V_{\medbullet\blacksquare}$ is $A(\theta,\phi)=\frac{1}{576}[37-12\cos2\theta-9\cos4\theta+48\sin2\theta\cos\phi+12(1+3\cos2\theta)\sin^2\theta\cos2\phi]$, 
which vanishes at $(\theta = -1/2\arcsin(2/3), \phi = 0)$, and is about two thirds of its maximal value at $(\theta = \pi/2, \phi = \pi/2)$. 
   
The atoms in the lattice representing  the duplicated logical qubits [`spheres' in Fig.~\ref{fig:f1}({\bf c})] and the ancilla (`cubes') are dressed with Rydberg $p$-states, $\ket{r_\medbullet}$ and $\ket{r_\blacksquare}$, respectively. The strong angular dependence of the Rydberg $p$-wave interaction allows for engineering the required interactions in Eq.~\eqref{eq:Code}. As shown in Fig.~\ref{fig:Rydberg}, the duplicated logical qubits are coupled along the $k$-axis, but are decoupled along the $i$-axis. Likewise, the ancilla are coupled only within each $k$-layer as required. 
The  interactions between the duplicated logical qubits and the ancilla,  as required to be programmable to encode the original Ising  spin couplings (see Eq.~\eqref{eq:JLR}),  are realized by the mixed interaction $V_{\medbullet\blacksquare}$ in Eq.~\eqref{eq:RydbergInt}.  
The programmability is achieved by considering spatial-resolved control over the Rydberg-dressing laser detuning~\cite{2010_Buchler_PRL,2010_Pupillo_PRL}. 
With the  Condon radius of Rydberg dressing set close to the lattice constant,  next-neighboring interactions are one-to-two orders of magnitude smaller than the nearest neighbor due to the $1/r^{6}$ decay of the van der Waals interaction~\cite{2010_Saffman_RMP}, which are thus negligible.

We remark here that the 3D quantum annealing architecture should be  slightly adjusted for Rydberg implementation  in order to suppress the unwanted couplings, for example the couplings between two quantum wire connectors with heights that differ by $1$. 
The direct couplings at height $h = 0$ should be replaced by introducing an additional quantum wire connector at height $h=1$. 
Each $k$-layer which contains the quantum wires labeled by $(ii')$ having  $\underline{i+i'} = k$, is splitt into two, according to the height of the quantum wire being even or odd.

With a proper choice of the detuning and Rabi frequency in the Rydberg dressing, an interaction strength at the order of several kHz can be obtained~\cite{glaetzle2014quantum}, which leads to a computation time of tens of milliseconds for problems studied in Sec.~\ref{sec:demo}. The computation time can be improved further by schedule optimization~\cite{2020_Lin_PRA,2020_Hsieh_arXiv}, or by adding counterdiabatic driving~\cite{2003_Demirplak_JPCA,2009_Berry_JPA,2017_Polkovnikov_PR}.  
Besides Rydberg dressing,  an alternative approach is to encode the qubit with one atomic ground state and one Rydberg state~\cite{browaeys2020many}, which has shorter lifetime but stronger interaction. 
Further considering F\"orster resonances controllable via external electric or microwave fields~\cite{weber2017calculation}, the interaction strength and the angular dependence would introduce a larger degree of tunability, making the Rydberg implementation a rather promising platform for programmable quantum annealing. 
All required ingredients in the Rydberg-based quantum annealing architecture are accessible with near-term quantum technology.

Another  experimental candidate for implementing the 3D quantum annealing architecture are polar molecules. By confining polar molecules in an array with optical tweezer techniques, the neighboring interactions can be made to order of kHz and are locally controllable~\cite{Anderegg2019AnOT,Yu_2019}. The natural long-range anisotropic interactions with polar molecules realize the fixed couplings in the quantum wire connectors and the long-range part of the interaction is expected to make the connectors more stiff, further suppressing defect excitations in the quantum wire. The programmable couplings can be achieved with AC Stark shift of tightly focused laser breams~\cite{Anderegg2019AnOT}.

\medskip       
\section{Discussion and Outlook} 

We have proposed Ising quantum wires to build long-range connectivity for programmable quantum annealing, by which a local quantum annealing architecture is developed. This setup can be embedded to a regular cubic lattice and only contains local fields and nearest-neighbor couplings. This can be implemented in experiments considering a system of ground state atoms in a spin dependent optical lattice, or Rydberg $p$-wave dressed atoms confined in a large-spacing optical lattice or tweezer arrays. 
The quantum annealing architecture  has reasonable robustness against finite temperature effects, 
and has good protection against readout errors from the logical qubit duplication.  
Our theory implies large-scale quantum annealing is accessible to near term optical lattice techniques. 
We have demonstrated how the theory applies to Max-Cut and prime factorization problems by simulating relatively small size problems on a classical computer. 
The present scheme of Ising quantum wires connecting spins in the quantum annealing problem can also be implemented in solid state architectures including superconducting devices~\cite{2019_Oliver_SCQubits}  and quantum dots~\cite{2013_Eriksson_RMP}, the main experimental challenge being the development of multilayered chips to represent the connecting wires. While the long term vision is an experimental realization of the 3D cubic architecture towards a programmable quantum annealer with all-to-all connectivity, we note that an experimental roadmap will first of all have to focus on basis building blocks like demonstration of m-port quantum wires and Ising interactions, which also provides interesting opportunities in quantum simulation of exotic spin models.

\medskip 
{\bf Acknowledgement.}
We acknowledge helpful discussion with Yu-Ao Chen, Saijun Wu, Han-Ning Dai, and Xingcan Yao. 
This work is supported by National Natural Science Foundation of China (Grants No. 11934002 and 11774067), National Program on Key Basic Research Project of China (Grant No. 2017YFA0304204), and Shanghai Municipal Science and Technology Major Project (Grant No. 2019SHZDZX01). 
X.Q. acknowledges support from National Postdoctoral Program for Innovative Talents of China under Grant No. BX20190083.
Work at Innsbruck is supported by the European Union program Horizon 2020 under Grants Agreement No. 817482 (PASQuanS).

\bibliography{ref}

\begin{thebibliography}{66}%
\makeatletter
\providecommand \@ifxundefined [1]{%
 \@ifx{#1\undefined}
}%
\providecommand \@ifnum [1]{%
 \ifnum #1\expandafter \@firstoftwo
 \else \expandafter \@secondoftwo
 \fi
}%
\providecommand \@ifx [1]{%
 \ifx #1\expandafter \@firstoftwo
 \else \expandafter \@secondoftwo
 \fi
}%
\providecommand \natexlab [1]{#1}%
\providecommand \enquote  [1]{``#1''}%
\providecommand \bibnamefont  [1]{#1}%
\providecommand \bibfnamefont [1]{#1}%
\providecommand \citenamefont [1]{#1}%
\providecommand \href@noop [0]{\@secondoftwo}%
\providecommand \href [0]{\begingroup \@sanitize@url \@href}%
\providecommand \@href[1]{\@@startlink{#1}\@@href}%
\providecommand \@@href[1]{\endgroup#1\@@endlink}%
\providecommand \@sanitize@url [0]{\catcode `\\12\catcode `\$12\catcode
  `\&12\catcode `\#12\catcode `\^12\catcode `\_12\catcode `\%12\relax}%
\providecommand \@@startlink[1]{}%
\providecommand \@@endlink[0]{}%
\providecommand \url  [0]{\begingroup\@sanitize@url \@url }%
\providecommand \@url [1]{\endgroup\@href {#1}{\urlprefix }}%
\providecommand \urlprefix  [0]{URL }%
\providecommand \Eprint [0]{\href }%
\providecommand \doibase [0]{http://dx.doi.org/}%
\providecommand \selectlanguage [0]{\@gobble}%
\providecommand \bibinfo  [0]{\@secondoftwo}%
\providecommand \bibfield  [0]{\@secondoftwo}%
\providecommand \translation [1]{[#1]}%
\providecommand \BibitemOpen [0]{}%
\providecommand \bibitemStop [0]{}%
\providecommand \bibitemNoStop [0]{.\EOS\space}%
\providecommand \EOS [0]{\spacefactor3000\relax}%
\providecommand \BibitemShut  [1]{\csname bibitem#1\endcsname}%
\let\auto@bib@innerbib\@empty
\bibitem [{\citenamefont {Farhi}\ \emph {et~al.}(2001)\citenamefont {Farhi},
  \citenamefont {Goldstone}, \citenamefont {Gutmann}, \citenamefont {Lapan},
  \citenamefont {Lundgren},\ and\ \citenamefont {Preda}}]{farhi_quantum_2000}%
  \BibitemOpen
  \bibfield  {author} {\bibinfo {author} {\bibfnamefont {E.}~\bibnamefont
  {Farhi}}, \bibinfo {author} {\bibfnamefont {J.}~\bibnamefont {Goldstone}},
  \bibinfo {author} {\bibfnamefont {S.}~\bibnamefont {Gutmann}}, \bibinfo
  {author} {\bibfnamefont {J.}~\bibnamefont {Lapan}}, \bibinfo {author}
  {\bibfnamefont {A.}~\bibnamefont {Lundgren}}, \ and\ \bibinfo {author}
  {\bibfnamefont {D.}~\bibnamefont {Preda}},\ }\href {\doibase
  10.1126/science.1057726} {\bibfield  {journal} {\bibinfo  {journal}
  {Science}\ }\textbf {\bibinfo {volume} {292}},\ \bibinfo {pages} {472}
  (\bibinfo {year} {2001})}\BibitemShut {NoStop}%
\bibitem [{\citenamefont {Das}\ and\ \citenamefont
  {Chakrabarti}(2008)}]{2008_Chakrabarti_RMP}%
  \BibitemOpen
  \bibfield  {author} {\bibinfo {author} {\bibfnamefont {A.}~\bibnamefont
  {Das}}\ and\ \bibinfo {author} {\bibfnamefont {B.~K.}\ \bibnamefont
  {Chakrabarti}},\ }\href {\doibase 10.1103/RevModPhys.80.1061} {\bibfield
  {journal} {\bibinfo  {journal} {Rev. Mod. Phys.}\ }\textbf {\bibinfo {volume}
  {80}},\ \bibinfo {pages} {1061} (\bibinfo {year} {2008})}\BibitemShut
  {NoStop}%
\bibitem [{\citenamefont {Albash}\ and\ \citenamefont
  {Lidar}(2018)}]{2018_Lidar_RMP}%
  \BibitemOpen
  \bibfield  {author} {\bibinfo {author} {\bibfnamefont {T.}~\bibnamefont
  {Albash}}\ and\ \bibinfo {author} {\bibfnamefont {D.~A.}\ \bibnamefont
  {Lidar}},\ }\href {\doibase 10.1103/RevModPhys.90.015002} {\bibfield
  {journal} {\bibinfo  {journal} {Rev. Mod. Phys.}\ }\textbf {\bibinfo {volume}
  {90}},\ \bibinfo {pages} {015002} (\bibinfo {year} {2018})}\BibitemShut
  {NoStop}%
\bibitem [{\citenamefont {Lucas}(2014)}]{2014_Lucas_FIP}%
  \BibitemOpen
  \bibfield  {author} {\bibinfo {author} {\bibfnamefont {A.}~\bibnamefont
  {Lucas}},\ }\href {\doibase 10.3389/fphy.2014.00005} {\bibfield  {journal}
  {\bibinfo  {journal} {Front. Phys.}\ }\textbf {\bibinfo {volume} {2}},\
  \bibinfo {pages} {5} (\bibinfo {year} {2014})}\BibitemShut {NoStop}%
\bibitem [{\citenamefont {J\"org}\ \emph {et~al.}(2008)\citenamefont {J\"org},
  \citenamefont {Krzakala}, \citenamefont {Kurchan},\ and\ \citenamefont
  {Maggs}}]{2008_Maggs_PRL}%
  \BibitemOpen
  \bibfield  {author} {\bibinfo {author} {\bibfnamefont {T.}~\bibnamefont
  {J\"org}}, \bibinfo {author} {\bibfnamefont {F.}~\bibnamefont {Krzakala}},
  \bibinfo {author} {\bibfnamefont {J.}~\bibnamefont {Kurchan}}, \ and\
  \bibinfo {author} {\bibfnamefont {A.~C.}\ \bibnamefont {Maggs}},\ }\href
  {\doibase 10.1103/PhysRevLett.101.147204} {\bibfield  {journal} {\bibinfo
  {journal} {Phys. Rev. Lett.}\ }\textbf {\bibinfo {volume} {101}},\ \bibinfo
  {pages} {147204} (\bibinfo {year} {2008})}\BibitemShut {NoStop}%
\bibitem [{\citenamefont {Dickson}\ and\ \citenamefont
  {Amin}(2011)}]{2011_Dickson_PRL}%
  \BibitemOpen
  \bibfield  {author} {\bibinfo {author} {\bibfnamefont {N.~G.}\ \bibnamefont
  {Dickson}}\ and\ \bibinfo {author} {\bibfnamefont {M.~H.~S.}\ \bibnamefont
  {Amin}},\ }\href {\doibase 10.1103/PhysRevLett.106.050502} {\bibfield
  {journal} {\bibinfo  {journal} {Phys. Rev. Lett.}\ }\textbf {\bibinfo
  {volume} {106}},\ \bibinfo {pages} {050502} (\bibinfo {year}
  {2011})}\BibitemShut {NoStop}%
\bibitem [{\citenamefont {Altshuler}\ \emph {et~al.}(2010)\citenamefont
  {Altshuler}, \citenamefont {Krovi},\ and\ \citenamefont
  {Roland}}]{Altshuler12446}%
  \BibitemOpen
  \bibfield  {author} {\bibinfo {author} {\bibfnamefont {B.}~\bibnamefont
  {Altshuler}}, \bibinfo {author} {\bibfnamefont {H.}~\bibnamefont {Krovi}}, \
  and\ \bibinfo {author} {\bibfnamefont {J.}~\bibnamefont {Roland}},\ }\href
  {\doibase 10.1073/pnas.1002116107} {\bibfield  {journal} {\bibinfo  {journal}
  {Proceedings of the National Academy of Sciences}\ }\textbf {\bibinfo
  {volume} {107}},\ \bibinfo {pages} {12446} (\bibinfo {year}
  {2010})}\BibitemShut {NoStop}%
\bibitem [{\citenamefont {Katzgraber}\ \emph {et~al.}(2014)\citenamefont
  {Katzgraber}, \citenamefont {Hamze},\ and\ \citenamefont
  {Andrist}}]{2014_Ruben_PRX}%
  \BibitemOpen
  \bibfield  {author} {\bibinfo {author} {\bibfnamefont {H.~G.}\ \bibnamefont
  {Katzgraber}}, \bibinfo {author} {\bibfnamefont {F.}~\bibnamefont {Hamze}}, \
  and\ \bibinfo {author} {\bibfnamefont {R.~S.}\ \bibnamefont {Andrist}},\
  }\href {\doibase 10.1103/PhysRevX.4.021008} {\bibfield  {journal} {\bibinfo
  {journal} {Phys. Rev. X}\ }\textbf {\bibinfo {volume} {4}},\ \bibinfo {pages}
  {021008} (\bibinfo {year} {2014})}\BibitemShut {NoStop}%
\bibitem [{\citenamefont {Ronnow}\ \emph {et~al.}(2014)\citenamefont {Ronnow},
  \citenamefont {Wang}, \citenamefont {Job}, \citenamefont {Boixo},
  \citenamefont {Isakov}, \citenamefont {Wecker}, \citenamefont {Martinis},
  \citenamefont {Lidar},\ and\ \citenamefont {Troyer}}]{Ronnow_2014}%
  \BibitemOpen
  \bibfield  {author} {\bibinfo {author} {\bibfnamefont {T.~F.}\ \bibnamefont
  {Ronnow}}, \bibinfo {author} {\bibfnamefont {Z.}~\bibnamefont {Wang}},
  \bibinfo {author} {\bibfnamefont {J.}~\bibnamefont {Job}}, \bibinfo {author}
  {\bibfnamefont {S.}~\bibnamefont {Boixo}}, \bibinfo {author} {\bibfnamefont
  {S.~V.}\ \bibnamefont {Isakov}}, \bibinfo {author} {\bibfnamefont
  {D.}~\bibnamefont {Wecker}}, \bibinfo {author} {\bibfnamefont {J.~M.}\
  \bibnamefont {Martinis}}, \bibinfo {author} {\bibfnamefont {D.~A.}\
  \bibnamefont {Lidar}}, \ and\ \bibinfo {author} {\bibfnamefont
  {M.}~\bibnamefont {Troyer}},\ }\href {\doibase 10.1126/science.1252319}
  {\bibfield  {journal} {\bibinfo  {journal} {Science}\ }\textbf {\bibinfo
  {volume} {345}},\ \bibinfo {pages} {420–424} (\bibinfo {year}
  {2014})}\BibitemShut {NoStop}%
\bibitem [{\citenamefont {Peng}\ \emph {et~al.}(2008)\citenamefont {Peng},
  \citenamefont {Liao}, \citenamefont {Xu}, \citenamefont {Qin}, \citenamefont
  {Zhou}, \citenamefont {Suter},\ and\ \citenamefont
  {Du}}]{2008_Du_Factorization}%
  \BibitemOpen
  \bibfield  {author} {\bibinfo {author} {\bibfnamefont {X.}~\bibnamefont
  {Peng}}, \bibinfo {author} {\bibfnamefont {Z.}~\bibnamefont {Liao}}, \bibinfo
  {author} {\bibfnamefont {N.}~\bibnamefont {Xu}}, \bibinfo {author}
  {\bibfnamefont {G.}~\bibnamefont {Qin}}, \bibinfo {author} {\bibfnamefont
  {X.}~\bibnamefont {Zhou}}, \bibinfo {author} {\bibfnamefont {D.}~\bibnamefont
  {Suter}}, \ and\ \bibinfo {author} {\bibfnamefont {J.}~\bibnamefont {Du}},\
  }\href {\doibase 10.1103/PhysRevLett.101.220405} {\bibfield  {journal}
  {\bibinfo  {journal} {Phys. Rev. Lett.}\ }\textbf {\bibinfo {volume} {101}},\
  \bibinfo {pages} {220405} (\bibinfo {year} {2008})}\BibitemShut {NoStop}%
\bibitem [{\citenamefont {Jiang}\ \emph {et~al.}(2018)\citenamefont {Jiang},
  \citenamefont {Britt}, \citenamefont {McCaskey}, \citenamefont {Humble},\
  and\ \citenamefont {Kais}}]{Kais_2018}%
  \BibitemOpen
  \bibfield  {author} {\bibinfo {author} {\bibfnamefont {S.}~\bibnamefont
  {Jiang}}, \bibinfo {author} {\bibfnamefont {K.~A.}\ \bibnamefont {Britt}},
  \bibinfo {author} {\bibfnamefont {A.~J.}\ \bibnamefont {McCaskey}}, \bibinfo
  {author} {\bibfnamefont {T.~S.}\ \bibnamefont {Humble}}, \ and\ \bibinfo
  {author} {\bibfnamefont {S.}~\bibnamefont {Kais}},\ }\href
  {https://doi.org/10.1038/s41598-018-36058-z} {\bibfield  {journal} {\bibinfo
  {journal} {Sci. Rep.}\ }\textbf {\bibinfo {volume} {8}},\ \bibinfo {pages}
  {17667} (\bibinfo {year} {2018})}\BibitemShut {NoStop}%
\bibitem [{\citenamefont {Zhou}\ \emph {et~al.}(2020)\citenamefont {Zhou},
  \citenamefont {Wang}, \citenamefont {Choi}, \citenamefont {Pichler},\ and\
  \citenamefont {Lukin}}]{zhou2018quantum}%
  \BibitemOpen
  \bibfield  {author} {\bibinfo {author} {\bibfnamefont {L.}~\bibnamefont
  {Zhou}}, \bibinfo {author} {\bibfnamefont {S.-T.}\ \bibnamefont {Wang}},
  \bibinfo {author} {\bibfnamefont {S.}~\bibnamefont {Choi}}, \bibinfo {author}
  {\bibfnamefont {H.}~\bibnamefont {Pichler}}, \ and\ \bibinfo {author}
  {\bibfnamefont {M.~D.}\ \bibnamefont {Lukin}},\ }\href {\doibase
  10.1103/PhysRevX.10.021067} {\bibfield  {journal} {\bibinfo  {journal} {Phys.
  Rev. X}\ }\textbf {\bibinfo {volume} {10}},\ \bibinfo {pages} {021067}
  (\bibinfo {year} {2020})}\BibitemShut {NoStop}%
\bibitem [{\citenamefont {King}\ \emph {et~al.}(2017)\citenamefont {King},
  \citenamefont {Yarkoni}, \citenamefont {Raymond}, \citenamefont {Ozfidan},
  \citenamefont {King}, \citenamefont {Nevisi}, \citenamefont {Hilton},\ and\
  \citenamefont {McGeoch}}]{king2017quantum}%
  \BibitemOpen
  \bibfield  {author} {\bibinfo {author} {\bibfnamefont {J.}~\bibnamefont
  {King}}, \bibinfo {author} {\bibfnamefont {S.}~\bibnamefont {Yarkoni}},
  \bibinfo {author} {\bibfnamefont {J.}~\bibnamefont {Raymond}}, \bibinfo
  {author} {\bibfnamefont {I.}~\bibnamefont {Ozfidan}}, \bibinfo {author}
  {\bibfnamefont {A.~D.}\ \bibnamefont {King}}, \bibinfo {author}
  {\bibfnamefont {M.~M.}\ \bibnamefont {Nevisi}}, \bibinfo {author}
  {\bibfnamefont {J.~P.}\ \bibnamefont {Hilton}}, \ and\ \bibinfo {author}
  {\bibfnamefont {C.~C.}\ \bibnamefont {McGeoch}},\ }\href
  {https://arxiv.org/abs/1701.04579} {\enquote {\bibinfo {title} {Quantum
  annealing amid local ruggedness and global frustration},}\ } (\bibinfo {year}
  {2017}),\ \Eprint {http://arxiv.org/abs/1701.04579} {arXiv:1701.04579
  [quant-ph]} \BibitemShut {NoStop}%
\bibitem [{\citenamefont {Boixo}\ \emph {et~al.}(2014)\citenamefont {Boixo},
  \citenamefont {Rønnow}, \citenamefont {Isakov}, \citenamefont {Wang},
  \citenamefont {Wecker}, \citenamefont {Lidar}, \citenamefont {Martinis},\
  and\ \citenamefont {Troyer}}]{2014_Troyer_NPhys}%
  \BibitemOpen
  \bibfield  {author} {\bibinfo {author} {\bibfnamefont {S.}~\bibnamefont
  {Boixo}}, \bibinfo {author} {\bibfnamefont {T.~F.}\ \bibnamefont {Rønnow}},
  \bibinfo {author} {\bibfnamefont {S.~V.}\ \bibnamefont {Isakov}}, \bibinfo
  {author} {\bibfnamefont {Z.}~\bibnamefont {Wang}}, \bibinfo {author}
  {\bibfnamefont {D.}~\bibnamefont {Wecker}}, \bibinfo {author} {\bibfnamefont
  {D.~A.}\ \bibnamefont {Lidar}}, \bibinfo {author} {\bibfnamefont {J.~M.}\
  \bibnamefont {Martinis}}, \ and\ \bibinfo {author} {\bibfnamefont
  {M.}~\bibnamefont {Troyer}},\ }\href {\doibase 10.1038/nphys2900} {\bibfield
  {journal} {\bibinfo  {journal} {Nat. Phys.}\ }\textbf {\bibinfo {volume}
  {10}},\ \bibinfo {pages} {218–} (\bibinfo {year} {2014})}\BibitemShut
  {NoStop}%
\bibitem [{\citenamefont {Lechner}\ \emph {et~al.}(2015)\citenamefont
  {Lechner}, \citenamefont {Hauke},\ and\ \citenamefont
  {Zoller}}]{Lechnere1500838}%
  \BibitemOpen
  \bibfield  {author} {\bibinfo {author} {\bibfnamefont {W.}~\bibnamefont
  {Lechner}}, \bibinfo {author} {\bibfnamefont {P.}~\bibnamefont {Hauke}}, \
  and\ \bibinfo {author} {\bibfnamefont {P.}~\bibnamefont {Zoller}},\ }\href
  {\doibase 10.1126/sciadv.1500838} {\bibfield  {journal} {\bibinfo  {journal}
  {Sci. Adv.}\ }\textbf {\bibinfo {volume} {1}},\ \bibinfo {pages} {e1500838}
  (\bibinfo {year} {2015})}\BibitemShut {NoStop}%
\bibitem [{\citenamefont {Rocchetto}\ \emph {et~al.}(2016)\citenamefont
  {Rocchetto}, \citenamefont {Benjamin},\ and\ \citenamefont
  {Li}}]{2016_Rocchetto_SA}%
  \BibitemOpen
  \bibfield  {author} {\bibinfo {author} {\bibfnamefont {A.}~\bibnamefont
  {Rocchetto}}, \bibinfo {author} {\bibfnamefont {S.~C.}\ \bibnamefont
  {Benjamin}}, \ and\ \bibinfo {author} {\bibfnamefont {Y.}~\bibnamefont
  {Li}},\ }\href {https://advances.sciencemag.org/content/2/10/e1601246}
  {\bibfield  {journal} {\bibinfo  {journal} {Sci. Adv.}\ }\textbf {\bibinfo
  {volume} {2}},\ \bibinfo {pages} {e1601246} (\bibinfo {year}
  {2016})}\BibitemShut {NoStop}%
\bibitem [{\citenamefont {Glaetzle}\ \emph {et~al.}(2017)\citenamefont
  {Glaetzle}, \citenamefont {van Bijnen}, \citenamefont {Zoller},\ and\
  \citenamefont {Lechner}}]{2017_Glaetzle_NC}%
  \BibitemOpen
  \bibfield  {author} {\bibinfo {author} {\bibfnamefont {A.~W.}\ \bibnamefont
  {Glaetzle}}, \bibinfo {author} {\bibfnamefont {R.~M.~W.}\ \bibnamefont {van
  Bijnen}}, \bibinfo {author} {\bibfnamefont {P.}~\bibnamefont {Zoller}}, \
  and\ \bibinfo {author} {\bibfnamefont {W.}~\bibnamefont {Lechner}},\ }\href
  {\doibase 10.1038/ncomms15813 (2017} {\bibfield  {journal} {\bibinfo
  {journal} {Nat. Commun.}\ }\textbf {\bibinfo {volume} {8}},\ \bibinfo {pages}
  {15813} (\bibinfo {year} {2017})}\BibitemShut {NoStop}%
\bibitem [{\citenamefont {Pichler}\ \emph {et~al.}(2018)\citenamefont
  {Pichler}, \citenamefont {Wang}, \citenamefont {Zhou}, \citenamefont {Choi},\
  and\ \citenamefont {Lukin}}]{pichler2018computational}%
  \BibitemOpen
  \bibfield  {author} {\bibinfo {author} {\bibfnamefont {H.}~\bibnamefont
  {Pichler}}, \bibinfo {author} {\bibfnamefont {S.-T.}\ \bibnamefont {Wang}},
  \bibinfo {author} {\bibfnamefont {L.}~\bibnamefont {Zhou}}, \bibinfo {author}
  {\bibfnamefont {S.}~\bibnamefont {Choi}}, \ and\ \bibinfo {author}
  {\bibfnamefont {M.~D.}\ \bibnamefont {Lukin}},\ }\href
  {https://arxiv.org/abs/1809.04954} {\enquote {\bibinfo {title} {Computational
  complexity of the {Rydberg} blockade in two dimensions},}\ } (\bibinfo {year}
  {2018}),\ \Eprint {http://arxiv.org/abs/1809.04954} {arXiv:1809.04954
  [quant-ph]} \BibitemShut {NoStop}%
\bibitem [{\citenamefont {Altman}\ \emph {et~al.}(2019)\citenamefont {Altman},
  \citenamefont {Brown}, \citenamefont {Carleo}, \citenamefont {Carr},
  \citenamefont {Demler}, \citenamefont {Chin}, \citenamefont {DeMarco},
  \citenamefont {Economou}, \citenamefont {Eriksson}, \citenamefont {Fu},
  \citenamefont {Greiner}, \citenamefont {Hazzard}, \citenamefont {Hulet},
  \citenamefont {Kollar}, \citenamefont {Lev}, \citenamefont {Lukin},
  \citenamefont {Ma}, \citenamefont {Mi}, \citenamefont {Misra}, \citenamefont
  {Monroe}, \citenamefont {Murch}, \citenamefont {Nazario}, \citenamefont {Ni},
  \citenamefont {Potter}, \citenamefont {Roushan}, \citenamefont {Saffman},
  \citenamefont {Schleier-Smith}, \citenamefont {Siddiqi}, \citenamefont
  {Simmonds}, \citenamefont {Singh}, \citenamefont {Spielman}, \citenamefont
  {Temme}, \citenamefont {Weiss}, \citenamefont {Vuckovic}, \citenamefont
  {Vuletic}, \citenamefont {Ye},\ and\ \citenamefont
  {Zwierlein}}]{altman2019quantum}%
  \BibitemOpen
  \bibfield  {author} {\bibinfo {author} {\bibfnamefont {E.}~\bibnamefont
  {Altman}}, \bibinfo {author} {\bibfnamefont {K.~R.}\ \bibnamefont {Brown}},
  \bibinfo {author} {\bibfnamefont {G.}~\bibnamefont {Carleo}}, \bibinfo
  {author} {\bibfnamefont {L.~D.}\ \bibnamefont {Carr}}, \bibinfo {author}
  {\bibfnamefont {E.}~\bibnamefont {Demler}}, \bibinfo {author} {\bibfnamefont
  {C.}~\bibnamefont {Chin}}, \bibinfo {author} {\bibfnamefont {B.}~\bibnamefont
  {DeMarco}}, \bibinfo {author} {\bibfnamefont {S.~E.}\ \bibnamefont
  {Economou}}, \bibinfo {author} {\bibfnamefont {M.~A.}\ \bibnamefont
  {Eriksson}}, \bibinfo {author} {\bibfnamefont {K.-M.~C.}\ \bibnamefont {Fu}},
  \bibinfo {author} {\bibfnamefont {M.}~\bibnamefont {Greiner}}, \bibinfo
  {author} {\bibfnamefont {K.~R.~A.}\ \bibnamefont {Hazzard}}, \bibinfo
  {author} {\bibfnamefont {R.~G.}\ \bibnamefont {Hulet}}, \bibinfo {author}
  {\bibfnamefont {A.~J.}\ \bibnamefont {Kollar}}, \bibinfo {author}
  {\bibfnamefont {B.~L.}\ \bibnamefont {Lev}}, \bibinfo {author} {\bibfnamefont
  {M.~D.}\ \bibnamefont {Lukin}}, \bibinfo {author} {\bibfnamefont
  {R.}~\bibnamefont {Ma}}, \bibinfo {author} {\bibfnamefont {X.}~\bibnamefont
  {Mi}}, \bibinfo {author} {\bibfnamefont {S.}~\bibnamefont {Misra}}, \bibinfo
  {author} {\bibfnamefont {C.}~\bibnamefont {Monroe}}, \bibinfo {author}
  {\bibfnamefont {K.}~\bibnamefont {Murch}}, \bibinfo {author} {\bibfnamefont
  {Z.}~\bibnamefont {Nazario}}, \bibinfo {author} {\bibfnamefont {K.-K.}\
  \bibnamefont {Ni}}, \bibinfo {author} {\bibfnamefont {A.~C.}\ \bibnamefont
  {Potter}}, \bibinfo {author} {\bibfnamefont {P.}~\bibnamefont {Roushan}},
  \bibinfo {author} {\bibfnamefont {M.}~\bibnamefont {Saffman}}, \bibinfo
  {author} {\bibfnamefont {M.}~\bibnamefont {Schleier-Smith}}, \bibinfo
  {author} {\bibfnamefont {I.}~\bibnamefont {Siddiqi}}, \bibinfo {author}
  {\bibfnamefont {R.}~\bibnamefont {Simmonds}}, \bibinfo {author}
  {\bibfnamefont {M.}~\bibnamefont {Singh}}, \bibinfo {author} {\bibfnamefont
  {I.~B.}\ \bibnamefont {Spielman}}, \bibinfo {author} {\bibfnamefont
  {K.}~\bibnamefont {Temme}}, \bibinfo {author} {\bibfnamefont {D.~S.}\
  \bibnamefont {Weiss}}, \bibinfo {author} {\bibfnamefont {J.}~\bibnamefont
  {Vuckovic}}, \bibinfo {author} {\bibfnamefont {V.}~\bibnamefont {Vuletic}},
  \bibinfo {author} {\bibfnamefont {J.}~\bibnamefont {Ye}}, \ and\ \bibinfo
  {author} {\bibfnamefont {M.}~\bibnamefont {Zwierlein}},\ }\href
  {https://arxiv.org/abs/1912.06938} {\enquote {\bibinfo {title} {Quantum
  simulators: Architectures and opportunities},}\ } (\bibinfo {year} {2019}),\
  \Eprint {http://arxiv.org/abs/1912.06938} {arXiv:1912.06938 [quant-ph]}
  \BibitemShut {NoStop}%
\bibitem [{\citenamefont {Alexeev}\ \emph {et~al.}(2019)\citenamefont
  {Alexeev}, \citenamefont {Bacon}, \citenamefont {Brown}, \citenamefont
  {Calderbank}, \citenamefont {Carr}, \citenamefont {Chong}, \citenamefont
  {DeMarco}, \citenamefont {Englund}, \citenamefont {Farhi}, \citenamefont
  {Fefferman}, \citenamefont {Gorshkov}, \citenamefont {Houck}, \citenamefont
  {Kim}, \citenamefont {Kimmel}, \citenamefont {Lange}, \citenamefont {Lloyd},
  \citenamefont {Lukin}, \citenamefont {Maslov}, \citenamefont {Maunz},
  \citenamefont {Monroe}, \citenamefont {Preskill}, \citenamefont {Roetteler},
  \citenamefont {Savage},\ and\ \citenamefont {Thompson}}]{alexeev2019quantum}%
  \BibitemOpen
  \bibfield  {author} {\bibinfo {author} {\bibfnamefont {Y.}~\bibnamefont
  {Alexeev}}, \bibinfo {author} {\bibfnamefont {D.}~\bibnamefont {Bacon}},
  \bibinfo {author} {\bibfnamefont {K.~R.}\ \bibnamefont {Brown}}, \bibinfo
  {author} {\bibfnamefont {R.}~\bibnamefont {Calderbank}}, \bibinfo {author}
  {\bibfnamefont {L.~D.}\ \bibnamefont {Carr}}, \bibinfo {author}
  {\bibfnamefont {F.~T.}\ \bibnamefont {Chong}}, \bibinfo {author}
  {\bibfnamefont {B.}~\bibnamefont {DeMarco}}, \bibinfo {author} {\bibfnamefont
  {D.}~\bibnamefont {Englund}}, \bibinfo {author} {\bibfnamefont
  {E.}~\bibnamefont {Farhi}}, \bibinfo {author} {\bibfnamefont
  {B.}~\bibnamefont {Fefferman}}, \bibinfo {author} {\bibfnamefont {A.~V.}\
  \bibnamefont {Gorshkov}}, \bibinfo {author} {\bibfnamefont {A.}~\bibnamefont
  {Houck}}, \bibinfo {author} {\bibfnamefont {J.}~\bibnamefont {Kim}}, \bibinfo
  {author} {\bibfnamefont {S.}~\bibnamefont {Kimmel}}, \bibinfo {author}
  {\bibfnamefont {M.}~\bibnamefont {Lange}}, \bibinfo {author} {\bibfnamefont
  {S.}~\bibnamefont {Lloyd}}, \bibinfo {author} {\bibfnamefont {M.~D.}\
  \bibnamefont {Lukin}}, \bibinfo {author} {\bibfnamefont {D.}~\bibnamefont
  {Maslov}}, \bibinfo {author} {\bibfnamefont {P.}~\bibnamefont {Maunz}},
  \bibinfo {author} {\bibfnamefont {C.}~\bibnamefont {Monroe}}, \bibinfo
  {author} {\bibfnamefont {J.}~\bibnamefont {Preskill}}, \bibinfo {author}
  {\bibfnamefont {M.}~\bibnamefont {Roetteler}}, \bibinfo {author}
  {\bibfnamefont {M.}~\bibnamefont {Savage}}, \ and\ \bibinfo {author}
  {\bibfnamefont {J.}~\bibnamefont {Thompson}},\ }\href
  {https://arxiv.org/abs/1912.07577} {\enquote {\bibinfo {title} {Quantum
  computer systems for scientific discovery},}\ } (\bibinfo {year} {2019}),\
  \Eprint {http://arxiv.org/abs/1912.07577} {arXiv:1912.07577 [quant-ph]}
  \BibitemShut {NoStop}%
\bibitem [{\citenamefont {Gross}\ and\ \citenamefont
  {Bloch}(2017)}]{gross2017quantum}%
  \BibitemOpen
  \bibfield  {author} {\bibinfo {author} {\bibfnamefont {C.}~\bibnamefont
  {Gross}}\ and\ \bibinfo {author} {\bibfnamefont {I.}~\bibnamefont {Bloch}},\
  }\href {\doibase 10.1126/science.aal3837} {\bibfield  {journal} {\bibinfo
  {journal} {Science}\ }\textbf {\bibinfo {volume} {357}},\ \bibinfo {pages}
  {995} (\bibinfo {year} {2017})}\BibitemShut {NoStop}%
\bibitem [{\citenamefont {Yi}\ \emph {et~al.}(2008)\citenamefont {Yi},
  \citenamefont {Daley}, \citenamefont {Pupillo},\ and\ \citenamefont
  {Zoller}}]{Yi_2008}%
  \BibitemOpen
  \bibfield  {author} {\bibinfo {author} {\bibfnamefont {W.}~\bibnamefont
  {Yi}}, \bibinfo {author} {\bibfnamefont {A.~J.}\ \bibnamefont {Daley}},
  \bibinfo {author} {\bibfnamefont {G.}~\bibnamefont {Pupillo}}, \ and\
  \bibinfo {author} {\bibfnamefont {P.}~\bibnamefont {Zoller}},\ }\href
  {\doibase 10.1088/1367-2630/10/7/073015} {\bibfield  {journal} {\bibinfo
  {journal} {New J. Phys.}\ }\textbf {\bibinfo {volume} {10}},\ \bibinfo
  {pages} {073015} (\bibinfo {year} {2008})}\BibitemShut {NoStop}%
\bibitem [{\citenamefont {Gauthier}\ \emph {et~al.}(2016)\citenamefont
  {Gauthier}, \citenamefont {Lenton}, \citenamefont {Parry}, \citenamefont
  {Baker}, \citenamefont {Davis}, \citenamefont {Rubinsztein-Dunlop},\ and\
  \citenamefont {Neely}}]{gauthier2016direct}%
  \BibitemOpen
  \bibfield  {author} {\bibinfo {author} {\bibfnamefont {G.}~\bibnamefont
  {Gauthier}}, \bibinfo {author} {\bibfnamefont {I.}~\bibnamefont {Lenton}},
  \bibinfo {author} {\bibfnamefont {N.~M.}\ \bibnamefont {Parry}}, \bibinfo
  {author} {\bibfnamefont {M.}~\bibnamefont {Baker}}, \bibinfo {author}
  {\bibfnamefont {M.}~\bibnamefont {Davis}}, \bibinfo {author} {\bibfnamefont
  {H.}~\bibnamefont {Rubinsztein-Dunlop}}, \ and\ \bibinfo {author}
  {\bibfnamefont {T.}~\bibnamefont {Neely}},\ }\href {\doibase
  10.1364/optica.3.001136} {\bibfield  {journal} {\bibinfo  {journal} {Optica}\
  }\textbf {\bibinfo {volume} {3}},\ \bibinfo {pages} {1136} (\bibinfo {year}
  {2016})}\BibitemShut {NoStop}%
\bibitem [{\citenamefont {Mazurenko}\ \emph {et~al.}(2017)\citenamefont
  {Mazurenko}, \citenamefont {Chiu}, \citenamefont {Ji}, \citenamefont
  {Parsons}, \citenamefont {Kan{\'a}sz-Nagy}, \citenamefont {Schmidt},
  \citenamefont {Grusdt}, \citenamefont {Demler}, \citenamefont {Greif},\ and\
  \citenamefont {Greiner}}]{mazurenko2017cold}%
  \BibitemOpen
  \bibfield  {author} {\bibinfo {author} {\bibfnamefont {A.}~\bibnamefont
  {Mazurenko}}, \bibinfo {author} {\bibfnamefont {C.~S.}\ \bibnamefont {Chiu}},
  \bibinfo {author} {\bibfnamefont {G.}~\bibnamefont {Ji}}, \bibinfo {author}
  {\bibfnamefont {M.~F.}\ \bibnamefont {Parsons}}, \bibinfo {author}
  {\bibfnamefont {M.}~\bibnamefont {Kan{\'a}sz-Nagy}}, \bibinfo {author}
  {\bibfnamefont {R.}~\bibnamefont {Schmidt}}, \bibinfo {author} {\bibfnamefont
  {F.}~\bibnamefont {Grusdt}}, \bibinfo {author} {\bibfnamefont
  {E.}~\bibnamefont {Demler}}, \bibinfo {author} {\bibfnamefont
  {D.}~\bibnamefont {Greif}}, \ and\ \bibinfo {author} {\bibfnamefont
  {M.}~\bibnamefont {Greiner}},\ }\href {\doibase 10.1038/nature22362}
  {\bibfield  {journal} {\bibinfo  {journal} {Nature}\ }\textbf {\bibinfo
  {volume} {545}},\ \bibinfo {pages} {462} (\bibinfo {year}
  {2017})}\BibitemShut {NoStop}%
\bibitem [{\citenamefont {McDonald}\ \emph {et~al.}(2019)\citenamefont
  {McDonald}, \citenamefont {Trisnadi}, \citenamefont {Yao},\ and\
  \citenamefont {Chin}}]{2019_Chin_PRX}%
  \BibitemOpen
  \bibfield  {author} {\bibinfo {author} {\bibfnamefont {M.}~\bibnamefont
  {McDonald}}, \bibinfo {author} {\bibfnamefont {J.}~\bibnamefont {Trisnadi}},
  \bibinfo {author} {\bibfnamefont {K.-X.}\ \bibnamefont {Yao}}, \ and\
  \bibinfo {author} {\bibfnamefont {C.}~\bibnamefont {Chin}},\ }\href {\doibase
  10.1103/PhysRevX.9.021001} {\bibfield  {journal} {\bibinfo  {journal} {Phys.
  Rev. X}\ }\textbf {\bibinfo {volume} {9}},\ \bibinfo {pages} {021001}
  (\bibinfo {year} {2019})}\BibitemShut {NoStop}%
\bibitem [{\citenamefont {Subhankar}\ \emph {et~al.}(2019)\citenamefont
  {Subhankar}, \citenamefont {Wang}, \citenamefont {Tsui}, \citenamefont
  {Rolston},\ and\ \citenamefont {Porto}}]{2019_Porto_PRX}%
  \BibitemOpen
  \bibfield  {author} {\bibinfo {author} {\bibfnamefont {S.}~\bibnamefont
  {Subhankar}}, \bibinfo {author} {\bibfnamefont {Y.}~\bibnamefont {Wang}},
  \bibinfo {author} {\bibfnamefont {T.-C.}\ \bibnamefont {Tsui}}, \bibinfo
  {author} {\bibfnamefont {S.~L.}\ \bibnamefont {Rolston}}, \ and\ \bibinfo
  {author} {\bibfnamefont {J.~V.}\ \bibnamefont {Porto}},\ }\href {\doibase
  10.1103/PhysRevX.9.021002} {\bibfield  {journal} {\bibinfo  {journal} {Phys.
  Rev. X}\ }\textbf {\bibinfo {volume} {9}},\ \bibinfo {pages} {021002}
  (\bibinfo {year} {2019})}\BibitemShut {NoStop}%
\bibitem [{\citenamefont {Qiu}\ \emph {et~al.}(2020)\citenamefont {Qiu},
  \citenamefont {Zou}, \citenamefont {Qi},\ and\ \citenamefont
  {Li}}]{qiu2020precise}%
  \BibitemOpen
  \bibfield  {author} {\bibinfo {author} {\bibfnamefont {X.}~\bibnamefont
  {Qiu}}, \bibinfo {author} {\bibfnamefont {J.}~\bibnamefont {Zou}}, \bibinfo
  {author} {\bibfnamefont {X.}~\bibnamefont {Qi}}, \ and\ \bibinfo {author}
  {\bibfnamefont {X.}~\bibnamefont {Li}},\ }\href {\doibase
  10.1038/s41534-020-00315-9} {\bibfield  {journal} {\bibinfo  {journal} {npj
  Quantum Information}\ }\textbf {\bibinfo {volume} {6}},\ \bibinfo {pages} {1}
  (\bibinfo {year} {2020})}\BibitemShut {NoStop}%
\bibitem [{\citenamefont {Honer}\ \emph {et~al.}(2010)\citenamefont {Honer},
  \citenamefont {Weimer}, \citenamefont {Pfau},\ and\ \citenamefont
  {B\"uchler}}]{2010_Buchler_PRL}%
  \BibitemOpen
  \bibfield  {author} {\bibinfo {author} {\bibfnamefont {J.}~\bibnamefont
  {Honer}}, \bibinfo {author} {\bibfnamefont {H.}~\bibnamefont {Weimer}},
  \bibinfo {author} {\bibfnamefont {T.}~\bibnamefont {Pfau}}, \ and\ \bibinfo
  {author} {\bibfnamefont {H.~P.}\ \bibnamefont {B\"uchler}},\ }\href {\doibase
  10.1103/PhysRevLett.105.160404} {\bibfield  {journal} {\bibinfo  {journal}
  {Phys. Rev. Lett.}\ }\textbf {\bibinfo {volume} {105}},\ \bibinfo {pages}
  {160404} (\bibinfo {year} {2010})}\BibitemShut {NoStop}%
\bibitem [{\citenamefont {Pupillo}\ \emph {et~al.}(2010)\citenamefont
  {Pupillo}, \citenamefont {Micheli}, \citenamefont {Boninsegni}, \citenamefont
  {Lesanovsky},\ and\ \citenamefont {Zoller}}]{2010_Pupillo_PRL}%
  \BibitemOpen
  \bibfield  {author} {\bibinfo {author} {\bibfnamefont {G.}~\bibnamefont
  {Pupillo}}, \bibinfo {author} {\bibfnamefont {A.}~\bibnamefont {Micheli}},
  \bibinfo {author} {\bibfnamefont {M.}~\bibnamefont {Boninsegni}}, \bibinfo
  {author} {\bibfnamefont {I.}~\bibnamefont {Lesanovsky}}, \ and\ \bibinfo
  {author} {\bibfnamefont {P.}~\bibnamefont {Zoller}},\ }\href {\doibase
  10.1103/PhysRevLett.104.223002} {\bibfield  {journal} {\bibinfo  {journal}
  {Phys. Rev. Lett.}\ }\textbf {\bibinfo {volume} {104}},\ \bibinfo {pages}
  {223002} (\bibinfo {year} {2010})}\BibitemShut {NoStop}%
\bibitem [{\citenamefont {Viteau}\ \emph {et~al.}(2011)\citenamefont {Viteau},
  \citenamefont {Bason}, \citenamefont {Radogostowicz}, \citenamefont
  {Malossi}, \citenamefont {Ciampini}, \citenamefont {Morsch},\ and\
  \citenamefont {Arimondo}}]{2011_Arimondo_PRL}%
  \BibitemOpen
  \bibfield  {author} {\bibinfo {author} {\bibfnamefont {M.}~\bibnamefont
  {Viteau}}, \bibinfo {author} {\bibfnamefont {M.~G.}\ \bibnamefont {Bason}},
  \bibinfo {author} {\bibfnamefont {J.}~\bibnamefont {Radogostowicz}}, \bibinfo
  {author} {\bibfnamefont {N.}~\bibnamefont {Malossi}}, \bibinfo {author}
  {\bibfnamefont {D.}~\bibnamefont {Ciampini}}, \bibinfo {author}
  {\bibfnamefont {O.}~\bibnamefont {Morsch}}, \ and\ \bibinfo {author}
  {\bibfnamefont {E.}~\bibnamefont {Arimondo}},\ }\href {\doibase
  10.1103/PhysRevLett.107.060402} {\bibfield  {journal} {\bibinfo  {journal}
  {Phys. Rev. Lett.}\ }\textbf {\bibinfo {volume} {107}},\ \bibinfo {pages}
  {060402} (\bibinfo {year} {2011})}\BibitemShut {NoStop}%
\bibitem [{\citenamefont {Wang}\ \emph {et~al.}(2016)\citenamefont {Wang},
  \citenamefont {Kumar}, \citenamefont {Wu},\ and\ \citenamefont
  {Weiss}}]{2016_Weiss_Science}%
  \BibitemOpen
  \bibfield  {author} {\bibinfo {author} {\bibfnamefont {Y.}~\bibnamefont
  {Wang}}, \bibinfo {author} {\bibfnamefont {A.}~\bibnamefont {Kumar}},
  \bibinfo {author} {\bibfnamefont {T.-Y.}\ \bibnamefont {Wu}}, \ and\ \bibinfo
  {author} {\bibfnamefont {D.~S.}\ \bibnamefont {Weiss}},\ }\href {\doibase
  10.1126/science.aaf2581} {\bibfield  {journal} {\bibinfo  {journal}
  {Science}\ }\textbf {\bibinfo {volume} {352}},\ \bibinfo {pages} {1562}
  (\bibinfo {year} {2016})}\BibitemShut {NoStop}%
\bibitem [{\citenamefont {Browaeys}\ and\ \citenamefont
  {Lahaye}(2020)}]{browaeys2020many}%
  \BibitemOpen
  \bibfield  {author} {\bibinfo {author} {\bibfnamefont {A.}~\bibnamefont
  {Browaeys}}\ and\ \bibinfo {author} {\bibfnamefont {T.}~\bibnamefont
  {Lahaye}},\ }\href {\doibase 10.1038/s41567-019-0733-z} {\bibfield  {journal}
  {\bibinfo  {journal} {Nat. Phys.}\ }\textbf {\bibinfo {volume} {16}},\
  \bibinfo {pages} {132} (\bibinfo {year} {2020})}\BibitemShut {NoStop}%
\bibitem [{\citenamefont {Duan}\ and\ \citenamefont
  {Monroe}(2010)}]{2010_Duan_RMP}%
  \BibitemOpen
  \bibfield  {author} {\bibinfo {author} {\bibfnamefont {L.-M.}\ \bibnamefont
  {Duan}}\ and\ \bibinfo {author} {\bibfnamefont {C.}~\bibnamefont {Monroe}},\
  }\href {\doibase 10.1103/RevModPhys.82.1209} {\bibfield  {journal} {\bibinfo
  {journal} {Rev. Mod. Phys.}\ }\textbf {\bibinfo {volume} {82}},\ \bibinfo
  {pages} {1209} (\bibinfo {year} {2010})}\BibitemShut {NoStop}%
\bibitem [{\citenamefont {Saffman}\ \emph {et~al.}(2010)\citenamefont
  {Saffman}, \citenamefont {Walker},\ and\ \citenamefont
  {M\o{}lmer}}]{2010_Saffman_RMP}%
  \BibitemOpen
  \bibfield  {author} {\bibinfo {author} {\bibfnamefont {M.}~\bibnamefont
  {Saffman}}, \bibinfo {author} {\bibfnamefont {T.~G.}\ \bibnamefont {Walker}},
  \ and\ \bibinfo {author} {\bibfnamefont {K.}~\bibnamefont {M\o{}lmer}},\
  }\href {\doibase 10.1103/RevModPhys.82.2313} {\bibfield  {journal} {\bibinfo
  {journal} {Rev. Mod. Phys.}\ }\textbf {\bibinfo {volume} {82}},\ \bibinfo
  {pages} {2313} (\bibinfo {year} {2010})}\BibitemShut {NoStop}%
\bibitem [{\citenamefont {Weinberg}(1995)}]{weinberg_1995}%
  \BibitemOpen
  \bibfield  {author} {\bibinfo {author} {\bibfnamefont {S.}~\bibnamefont
  {Weinberg}},\ }\href {\doibase 10.1017/CBO9781139644167} {\emph {\bibinfo
  {title} {The Quantum Theory of Fields}}},\ Vol.~\bibinfo {volume} {I}\
  (\bibinfo  {publisher} {Cambridge University Press, Cambridge},\ \bibinfo
  {year} {1995})\BibitemShut {NoStop}%
\bibitem [{\citenamefont {Clark}\ \emph {et~al.}(1990)\citenamefont {Clark},
  \citenamefont {Colbourn},\ and\ \citenamefont
  {Johnson}}]{1990_Clark_UDgraphs}%
  \BibitemOpen
  \bibfield  {author} {\bibinfo {author} {\bibfnamefont {B.}~\bibnamefont
  {Clark}}, \bibinfo {author} {\bibfnamefont {C.}~\bibnamefont {Colbourn}}, \
  and\ \bibinfo {author} {\bibfnamefont {D.}~\bibnamefont {Johnson}},\ }\href
  {\doibase 10.1016/0012-365X(90)90358-O} {\bibfield  {journal} {\bibinfo
  {journal} {Discrete Math.}\ }\textbf {\bibinfo {volume} {86}},\ \bibinfo
  {pages} {165} (\bibinfo {year} {1990})}\BibitemShut {NoStop}%
\bibitem [{\citenamefont {Crawford}\ and\ \citenamefont
  {Auton}(1993)}]{1993_Crawfor_AAAI}%
  \BibitemOpen
  \bibfield  {author} {\bibinfo {author} {\bibfnamefont {J.~M.}\ \bibnamefont
  {Crawford}}\ and\ \bibinfo {author} {\bibfnamefont {L.~D.}\ \bibnamefont
  {Auton}},\ }in\ \href {http://www.aaai.org/Library/AAAI/1993/aaai93-004.php}
  {\emph {\bibinfo {booktitle} {Proceedings of the 11th National Conference on
  Artificial Intelligence}}},\ Vol.~\bibinfo {volume} {93}\ (\bibinfo
  {publisher} {AAAI, Menlo Park},\ \bibinfo {year} {1993})\ pp.\ \bibinfo
  {pages} {21--27}\BibitemShut {NoStop}%
\bibitem [{\citenamefont {Barab{\'a}si}\ and\ \citenamefont
  {Albert}(1999)}]{Barabasi}%
  \BibitemOpen
  \bibfield  {author} {\bibinfo {author} {\bibfnamefont {A.-L.}\ \bibnamefont
  {Barab{\'a}si}}\ and\ \bibinfo {author} {\bibfnamefont {R.}~\bibnamefont
  {Albert}},\ }\href {\doibase 10.1126/science.286.5439.509} {\bibfield
  {journal} {\bibinfo  {journal} {Science}\ }\textbf {\bibinfo {volume}
  {286}},\ \bibinfo {pages} {509} (\bibinfo {year} {1999})}\BibitemShut
  {NoStop}%
\bibitem [{\citenamefont {Kjaergaard}\ \emph {et~al.}(2020)\citenamefont
  {Kjaergaard}, \citenamefont {Schwartz}, \citenamefont {Braumüller},
  \citenamefont {Krantz}, \citenamefont {Wang}, \citenamefont {Gustavsson},\
  and\ \citenamefont {Oliver}}]{2019_Oliver_SCQubits}%
  \BibitemOpen
  \bibfield  {author} {\bibinfo {author} {\bibfnamefont {M.}~\bibnamefont
  {Kjaergaard}}, \bibinfo {author} {\bibfnamefont {M.~E.}\ \bibnamefont
  {Schwartz}}, \bibinfo {author} {\bibfnamefont {J.}~\bibnamefont
  {Braumüller}}, \bibinfo {author} {\bibfnamefont {P.}~\bibnamefont {Krantz}},
  \bibinfo {author} {\bibfnamefont {J.~I.-J.}\ \bibnamefont {Wang}}, \bibinfo
  {author} {\bibfnamefont {S.}~\bibnamefont {Gustavsson}}, \ and\ \bibinfo
  {author} {\bibfnamefont {W.~D.}\ \bibnamefont {Oliver}},\ }\href {\doibase
  10.1146/annurev-conmatphys-031119-050605} {\bibfield  {journal} {\bibinfo
  {journal} {Annual Review of Condensed Matter Physics}\ }\textbf {\bibinfo
  {volume} {11}},\ \bibinfo {pages} {369} (\bibinfo {year} {2020})}\BibitemShut
  {NoStop}%
\bibitem [{\citenamefont {Zwanenburg}\ \emph {et~al.}(2013)\citenamefont
  {Zwanenburg}, \citenamefont {Dzurak}, \citenamefont {Morello}, \citenamefont
  {Simmons}, \citenamefont {Hollenberg}, \citenamefont {Klimeck}, \citenamefont
  {Rogge}, \citenamefont {Coppersmith},\ and\ \citenamefont
  {Eriksson}}]{2013_Eriksson_RMP}%
  \BibitemOpen
  \bibfield  {author} {\bibinfo {author} {\bibfnamefont {F.~A.}\ \bibnamefont
  {Zwanenburg}}, \bibinfo {author} {\bibfnamefont {A.~S.}\ \bibnamefont
  {Dzurak}}, \bibinfo {author} {\bibfnamefont {A.}~\bibnamefont {Morello}},
  \bibinfo {author} {\bibfnamefont {M.~Y.}\ \bibnamefont {Simmons}}, \bibinfo
  {author} {\bibfnamefont {L.~C.~L.}\ \bibnamefont {Hollenberg}}, \bibinfo
  {author} {\bibfnamefont {G.}~\bibnamefont {Klimeck}}, \bibinfo {author}
  {\bibfnamefont {S.}~\bibnamefont {Rogge}}, \bibinfo {author} {\bibfnamefont
  {S.~N.}\ \bibnamefont {Coppersmith}}, \ and\ \bibinfo {author} {\bibfnamefont
  {M.~A.}\ \bibnamefont {Eriksson}},\ }\href {\doibase
  10.1103/RevModPhys.85.961} {\bibfield  {journal} {\bibinfo  {journal} {Rev.
  Mod. Phys.}\ }\textbf {\bibinfo {volume} {85}},\ \bibinfo {pages} {961}
  (\bibinfo {year} {2013})}\BibitemShut {NoStop}%
\bibitem [{\citenamefont {Lin}\ \emph {et~al.}(2020)\citenamefont {Lin},
  \citenamefont {Lai},\ and\ \citenamefont {Li}}]{2020_Lin_PRA}%
  \BibitemOpen
  \bibfield  {author} {\bibinfo {author} {\bibfnamefont {J.}~\bibnamefont
  {Lin}}, \bibinfo {author} {\bibfnamefont {Z.~Y.}\ \bibnamefont {Lai}}, \ and\
  \bibinfo {author} {\bibfnamefont {X.}~\bibnamefont {Li}},\ }\href {\doibase
  10.1103/PhysRevA.101.052327} {\bibfield  {journal} {\bibinfo  {journal}
  {Phys. Rev. A}\ }\textbf {\bibinfo {volume} {101}},\ \bibinfo {pages}
  {052327} (\bibinfo {year} {2020})}\BibitemShut {NoStop}%
\bibitem [{\citenamefont {Boros}\ and\ \citenamefont
  {Hammer}(2002)}]{2002_Hammer_Math}%
  \BibitemOpen
  \bibfield  {author} {\bibinfo {author} {\bibfnamefont {E.}~\bibnamefont
  {Boros}}\ and\ \bibinfo {author} {\bibfnamefont {P.~L.}\ \bibnamefont
  {Hammer}},\ }\href {\doibase 10.1016/S0166-218X(01)00341-9} {\bibfield
  {journal} {\bibinfo  {journal} {Discrete Appl. Math.}\ }\textbf {\bibinfo
  {volume} {123}},\ \bibinfo {pages} {155} (\bibinfo {year}
  {2002})}\BibitemShut {NoStop}%
\bibitem [{\citenamefont {Altland}\ and\ \citenamefont
  {Simons}(2010)}]{Altland-Simons}%
  \BibitemOpen
  \bibfield  {author} {\bibinfo {author} {\bibfnamefont {A.}~\bibnamefont
  {Altland}}\ and\ \bibinfo {author} {\bibfnamefont {B.~D.}\ \bibnamefont
  {Simons}},\ }\href {\doibase 10.1017/CBO9780511789984} {\emph {\bibinfo
  {title} {Condensed matter field theory}}}\ (\bibinfo  {publisher} {Cambridge
  university press, New York},\ \bibinfo {year} {2010})\BibitemShut {NoStop}%
\bibitem [{\citenamefont {Kibble}(1976)}]{Kibble_1976}%
  \BibitemOpen
  \bibfield  {author} {\bibinfo {author} {\bibfnamefont {T.~W.~B.}\
  \bibnamefont {Kibble}},\ }\href {\doibase 10.1088/0305-4470/9/8/029}
  {\bibfield  {journal} {\bibinfo  {journal} {J. Phys. A: Math. Gen.}\ }\textbf
  {\bibinfo {volume} {9}},\ \bibinfo {pages} {1387} (\bibinfo {year}
  {1976})}\BibitemShut {NoStop}%
\bibitem [{\citenamefont {Zurek}(1985)}]{1985_Zurek}%
  \BibitemOpen
  \bibfield  {author} {\bibinfo {author} {\bibfnamefont {W.~H.}\ \bibnamefont
  {Zurek}},\ }\href {https://www.nature.com/articles/317505a0} {\bibfield
  {journal} {\bibinfo  {journal} {Nature}\ }\textbf {\bibinfo {volume} {317}},\
  \bibinfo {pages} {505} (\bibinfo {year} {1985})}\BibitemShut {NoStop}%
\bibitem [{\citenamefont {Chen}\ \emph {et~al.}(2020)\citenamefont {Chen},
  \citenamefont {Chen}, \citenamefont {Lee}, \citenamefont {Zhang},\ and\
  \citenamefont {Hsieh}}]{2020_Hsieh_arXiv}%
  \BibitemOpen
  \bibfield  {author} {\bibinfo {author} {\bibfnamefont {Y.-Q.}\ \bibnamefont
  {Chen}}, \bibinfo {author} {\bibfnamefont {Y.}~\bibnamefont {Chen}}, \bibinfo
  {author} {\bibfnamefont {C.-K.}\ \bibnamefont {Lee}}, \bibinfo {author}
  {\bibfnamefont {S.}~\bibnamefont {Zhang}}, \ and\ \bibinfo {author}
  {\bibfnamefont {C.-Y.}\ \bibnamefont {Hsieh}},\ }\href
  {https://arxiv.org/abs/2004.02836} {\enquote {\bibinfo {title} {Optimizing
  quantum annealing schedules: From {Monte Carlo} tree search to
  {QuantumZero}},}\ } (\bibinfo {year} {2020}),\ \Eprint
  {http://arxiv.org/abs/2004.02836} {arXiv:2004.02836 [quant-ph]} \BibitemShut
  {NoStop}%
\bibitem [{\citenamefont {Demirplak}\ and\ \citenamefont
  {Rice}(2003)}]{2003_Demirplak_JPCA}%
  \BibitemOpen
  \bibfield  {author} {\bibinfo {author} {\bibfnamefont {M.}~\bibnamefont
  {Demirplak}}\ and\ \bibinfo {author} {\bibfnamefont {S.~A.}\ \bibnamefont
  {Rice}},\ }\href {https://pubs.acs.org/doi/abs/10.1021/jp030708a} {\bibfield
  {journal} {\bibinfo  {journal} {The Journal of Physical Chemistry A}\
  }\textbf {\bibinfo {volume} {107}},\ \bibinfo {pages} {9937} (\bibinfo {year}
  {2003})}\BibitemShut {NoStop}%
\bibitem [{\citenamefont {Berry}(2009)}]{2009_Berry_JPA}%
  \BibitemOpen
  \bibfield  {author} {\bibinfo {author} {\bibfnamefont {M.~V.}\ \bibnamefont
  {Berry}},\ }\href
  {https://iopscience.iop.org/article/10.1088/1751-8113/42/36/365303/meta}
  {\bibfield  {journal} {\bibinfo  {journal} {Journal of Physics A:
  Mathematical and Theoretical}\ }\textbf {\bibinfo {volume} {42}},\ \bibinfo
  {pages} {365303} (\bibinfo {year} {2009})}\BibitemShut {NoStop}%
\bibitem [{\citenamefont {Kolodrubetz}\ \emph {et~al.}(2017)\citenamefont
  {Kolodrubetz}, \citenamefont {Sels}, \citenamefont {Mehta},\ and\
  \citenamefont {Polkovnikov}}]{2017_Polkovnikov_PR}%
  \BibitemOpen
  \bibfield  {author} {\bibinfo {author} {\bibfnamefont {M.}~\bibnamefont
  {Kolodrubetz}}, \bibinfo {author} {\bibfnamefont {D.}~\bibnamefont {Sels}},
  \bibinfo {author} {\bibfnamefont {P.}~\bibnamefont {Mehta}}, \ and\ \bibinfo
  {author} {\bibfnamefont {A.}~\bibnamefont {Polkovnikov}},\ }\href
  {https://www.sciencedirect.com/science/article/abs/pii/S0370157317301989}
  {\bibfield  {journal} {\bibinfo  {journal} {Phys. Rep.}\ }\textbf {\bibinfo
  {volume} {697}},\ \bibinfo {pages} {1} (\bibinfo {year} {2017})}\BibitemShut
  {NoStop}%
\bibitem [{\citenamefont {Duan}\ \emph {et~al.}(2003)\citenamefont {Duan},
  \citenamefont {Demler},\ and\ \citenamefont {Lukin}}]{2003_Duan_PRL}%
  \BibitemOpen
  \bibfield  {author} {\bibinfo {author} {\bibfnamefont {L.-M.}\ \bibnamefont
  {Duan}}, \bibinfo {author} {\bibfnamefont {E.}~\bibnamefont {Demler}}, \ and\
  \bibinfo {author} {\bibfnamefont {M.~D.}\ \bibnamefont {Lukin}},\ }\href
  {\doibase 10.1103/PhysRevLett.91.090402} {\bibfield  {journal} {\bibinfo
  {journal} {Phys. Rev. Lett.}\ }\textbf {\bibinfo {volume} {91}},\ \bibinfo
  {pages} {090402} (\bibinfo {year} {2003})}\BibitemShut {NoStop}%
\bibitem [{\citenamefont {Jaksch}\ \emph {et~al.}(1999)\citenamefont {Jaksch},
  \citenamefont {Briegel}, \citenamefont {Cirac}, \citenamefont {Gardiner},\
  and\ \citenamefont {Zoller}}]{1999_Jaksch_PRL}%
  \BibitemOpen
  \bibfield  {author} {\bibinfo {author} {\bibfnamefont {D.}~\bibnamefont
  {Jaksch}}, \bibinfo {author} {\bibfnamefont {H.-J.}\ \bibnamefont {Briegel}},
  \bibinfo {author} {\bibfnamefont {J.~I.}\ \bibnamefont {Cirac}}, \bibinfo
  {author} {\bibfnamefont {C.~W.}\ \bibnamefont {Gardiner}}, \ and\ \bibinfo
  {author} {\bibfnamefont {P.}~\bibnamefont {Zoller}},\ }\href {\doibase
  10.1103/PhysRevLett.82.1975} {\bibfield  {journal} {\bibinfo  {journal}
  {Phys. Rev. Lett.}\ }\textbf {\bibinfo {volume} {82}},\ \bibinfo {pages}
  {1975} (\bibinfo {year} {1999})}\BibitemShut {NoStop}%
\bibitem [{\citenamefont {Liu}\ \emph {et~al.}(2004)\citenamefont {Liu},
  \citenamefont {Wilczek},\ and\ \citenamefont {Zoller}}]{2004_Vincent_PRA}%
  \BibitemOpen
  \bibfield  {author} {\bibinfo {author} {\bibfnamefont {W.~V.}\ \bibnamefont
  {Liu}}, \bibinfo {author} {\bibfnamefont {F.}~\bibnamefont {Wilczek}}, \ and\
  \bibinfo {author} {\bibfnamefont {P.}~\bibnamefont {Zoller}},\ }\href
  {\doibase 10.1103/PhysRevA.70.033603} {\bibfield  {journal} {\bibinfo
  {journal} {Phys. Rev. A}\ }\textbf {\bibinfo {volume} {70}},\ \bibinfo
  {pages} {033603} (\bibinfo {year} {2004})}\BibitemShut {NoStop}%
\bibitem [{\citenamefont {Lee}\ \emph {et~al.}(2007)\citenamefont {Lee},
  \citenamefont {Anderlini}, \citenamefont {Brown}, \citenamefont
  {Sebby-Strabley}, \citenamefont {Phillips},\ and\ \citenamefont
  {Porto}}]{2007_Porto_PRL}%
  \BibitemOpen
  \bibfield  {author} {\bibinfo {author} {\bibfnamefont {P.~J.}\ \bibnamefont
  {Lee}}, \bibinfo {author} {\bibfnamefont {M.}~\bibnamefont {Anderlini}},
  \bibinfo {author} {\bibfnamefont {B.~L.}\ \bibnamefont {Brown}}, \bibinfo
  {author} {\bibfnamefont {J.}~\bibnamefont {Sebby-Strabley}}, \bibinfo
  {author} {\bibfnamefont {W.~D.}\ \bibnamefont {Phillips}}, \ and\ \bibinfo
  {author} {\bibfnamefont {J.~V.}\ \bibnamefont {Porto}},\ }\href {\doibase
  10.1103/PhysRevLett.99.020402} {\bibfield  {journal} {\bibinfo  {journal}
  {Phys. Rev. Lett.}\ }\textbf {\bibinfo {volume} {99}},\ \bibinfo {pages}
  {020402} (\bibinfo {year} {2007})}\BibitemShut {NoStop}%
\bibitem [{\citenamefont {Daley}\ \emph {et~al.}(2008)\citenamefont {Daley},
  \citenamefont {Boyd}, \citenamefont {Ye},\ and\ \citenamefont
  {Zoller}}]{2008_Zoller_PRL}%
  \BibitemOpen
  \bibfield  {author} {\bibinfo {author} {\bibfnamefont {A.~J.}\ \bibnamefont
  {Daley}}, \bibinfo {author} {\bibfnamefont {M.~M.}\ \bibnamefont {Boyd}},
  \bibinfo {author} {\bibfnamefont {J.}~\bibnamefont {Ye}}, \ and\ \bibinfo
  {author} {\bibfnamefont {P.}~\bibnamefont {Zoller}},\ }\href {\doibase
  10.1103/PhysRevLett.101.170504} {\bibfield  {journal} {\bibinfo  {journal}
  {Phys. Rev. Lett.}\ }\textbf {\bibinfo {volume} {101}},\ \bibinfo {pages}
  {170504} (\bibinfo {year} {2008})}\BibitemShut {NoStop}%
\bibitem [{\citenamefont {Dai}\ \emph {et~al.}(2017)\citenamefont {Dai},
  \citenamefont {Yang}, \citenamefont {Reingruber}, \citenamefont {Sun},
  \citenamefont {Xu}, \citenamefont {Chen}, \citenamefont {Yuan},\ and\
  \citenamefont {Pan}}]{2017_Pan_ToricCode}%
  \BibitemOpen
  \bibfield  {author} {\bibinfo {author} {\bibfnamefont {H.-N.}\ \bibnamefont
  {Dai}}, \bibinfo {author} {\bibfnamefont {B.}~\bibnamefont {Yang}}, \bibinfo
  {author} {\bibfnamefont {A.}~\bibnamefont {Reingruber}}, \bibinfo {author}
  {\bibfnamefont {H.}~\bibnamefont {Sun}}, \bibinfo {author} {\bibfnamefont
  {X.-F.}\ \bibnamefont {Xu}}, \bibinfo {author} {\bibfnamefont {Y.-A.}\
  \bibnamefont {Chen}}, \bibinfo {author} {\bibfnamefont {Z.-S.}\ \bibnamefont
  {Yuan}}, \ and\ \bibinfo {author} {\bibfnamefont {J.-W.}\ \bibnamefont
  {Pan}},\ }\href {\doibase 10.1038/nphys4243} {\bibfield  {journal} {\bibinfo
  {journal} {Nat. Phys.}\ }\textbf {\bibinfo {volume} {13}},\ \bibinfo {pages}
  {1195} (\bibinfo {year} {2017})}\BibitemShut {NoStop}%
\bibitem [{\citenamefont {Chin}\ \emph {et~al.}(2010)\citenamefont {Chin},
  \citenamefont {Grimm}, \citenamefont {Julienne},\ and\ \citenamefont
  {Tiesinga}}]{2010_Chin_RMP}%
  \BibitemOpen
  \bibfield  {author} {\bibinfo {author} {\bibfnamefont {C.}~\bibnamefont
  {Chin}}, \bibinfo {author} {\bibfnamefont {R.}~\bibnamefont {Grimm}},
  \bibinfo {author} {\bibfnamefont {P.}~\bibnamefont {Julienne}}, \ and\
  \bibinfo {author} {\bibfnamefont {E.}~\bibnamefont {Tiesinga}},\ }\href
  {\doibase 10.1103/RevModPhys.82.1225} {\bibfield  {journal} {\bibinfo
  {journal} {Rev. Mod. Phys.}\ }\textbf {\bibinfo {volume} {82}},\ \bibinfo
  {pages} {1225} (\bibinfo {year} {2010})}\BibitemShut {NoStop}%
\bibitem [{\citenamefont {Mamaev}\ \emph {et~al.}(2019)\citenamefont {Mamaev},
  \citenamefont {Blatt}, \citenamefont {Ye},\ and\ \citenamefont
  {Rey}}]{2019_Rey_PRL}%
  \BibitemOpen
  \bibfield  {author} {\bibinfo {author} {\bibfnamefont {M.}~\bibnamefont
  {Mamaev}}, \bibinfo {author} {\bibfnamefont {R.}~\bibnamefont {Blatt}},
  \bibinfo {author} {\bibfnamefont {J.}~\bibnamefont {Ye}}, \ and\ \bibinfo
  {author} {\bibfnamefont {A.~M.}\ \bibnamefont {Rey}},\ }\href
  {https://journals.aps.org/prl/abstract/10.1103/PhysRevLett.122.160402}
  {\bibfield  {journal} {\bibinfo  {journal} {Phys. Rev. Lett.}\ }\textbf
  {\bibinfo {volume} {122}},\ \bibinfo {pages} {160402} (\bibinfo {year}
  {2019})}\BibitemShut {NoStop}%
\bibitem [{\citenamefont {Zhang}\ \emph {et~al.}(2015)\citenamefont {Zhang},
  \citenamefont {Cheng}, \citenamefont {Zhai},\ and\ \citenamefont
  {Zhang}}]{2015_Zhang_PRL}%
  \BibitemOpen
  \bibfield  {author} {\bibinfo {author} {\bibfnamefont {R.}~\bibnamefont
  {Zhang}}, \bibinfo {author} {\bibfnamefont {Y.}~\bibnamefont {Cheng}},
  \bibinfo {author} {\bibfnamefont {H.}~\bibnamefont {Zhai}}, \ and\ \bibinfo
  {author} {\bibfnamefont {P.}~\bibnamefont {Zhang}},\ }\href {\doibase
  10.1103/PhysRevLett.115.135301} {\bibfield  {journal} {\bibinfo  {journal}
  {Phys. Rev. Lett.}\ }\textbf {\bibinfo {volume} {115}},\ \bibinfo {pages}
  {135301} (\bibinfo {year} {2015})}\BibitemShut {NoStop}%
\bibitem [{\citenamefont {H\"ofer}\ \emph {et~al.}(2015)\citenamefont
  {H\"ofer}, \citenamefont {Riegger}, \citenamefont {Scazza}, \citenamefont
  {Hofrichter}, \citenamefont {Fernandes}, \citenamefont {Parish},
  \citenamefont {Levinsen}, \citenamefont {Bloch},\ and\ \citenamefont
  {F\"olling}}]{2015_Folling_PRL}%
  \BibitemOpen
  \bibfield  {author} {\bibinfo {author} {\bibfnamefont {M.}~\bibnamefont
  {H\"ofer}}, \bibinfo {author} {\bibfnamefont {L.}~\bibnamefont {Riegger}},
  \bibinfo {author} {\bibfnamefont {F.}~\bibnamefont {Scazza}}, \bibinfo
  {author} {\bibfnamefont {C.}~\bibnamefont {Hofrichter}}, \bibinfo {author}
  {\bibfnamefont {D.~R.}\ \bibnamefont {Fernandes}}, \bibinfo {author}
  {\bibfnamefont {M.~M.}\ \bibnamefont {Parish}}, \bibinfo {author}
  {\bibfnamefont {J.}~\bibnamefont {Levinsen}}, \bibinfo {author}
  {\bibfnamefont {I.}~\bibnamefont {Bloch}}, \ and\ \bibinfo {author}
  {\bibfnamefont {S.}~\bibnamefont {F\"olling}},\ }\href {\doibase
  10.1103/PhysRevLett.115.265302} {\bibfield  {journal} {\bibinfo  {journal}
  {Phys. Rev. Lett.}\ }\textbf {\bibinfo {volume} {115}},\ \bibinfo {pages}
  {265302} (\bibinfo {year} {2015})}\BibitemShut {NoStop}%
\bibitem [{\citenamefont {Kolkowitz}\ \emph {et~al.}(2017)\citenamefont
  {Kolkowitz}, \citenamefont {Bromley}, \citenamefont {Bothwell}, \citenamefont
  {Wall}, \citenamefont {Marti}, \citenamefont {Koller}, \citenamefont {Zhang},
  \citenamefont {Rey},\ and\ \citenamefont {Ye}}]{2017_Ye_Kolkowitz_Nature}%
  \BibitemOpen
  \bibfield  {author} {\bibinfo {author} {\bibfnamefont {S.}~\bibnamefont
  {Kolkowitz}}, \bibinfo {author} {\bibfnamefont {S.}~\bibnamefont {Bromley}},
  \bibinfo {author} {\bibfnamefont {T.}~\bibnamefont {Bothwell}}, \bibinfo
  {author} {\bibfnamefont {M.}~\bibnamefont {Wall}}, \bibinfo {author}
  {\bibfnamefont {G.}~\bibnamefont {Marti}}, \bibinfo {author} {\bibfnamefont
  {A.}~\bibnamefont {Koller}}, \bibinfo {author} {\bibfnamefont
  {X.}~\bibnamefont {Zhang}}, \bibinfo {author} {\bibfnamefont
  {A.}~\bibnamefont {Rey}}, \ and\ \bibinfo {author} {\bibfnamefont
  {J.}~\bibnamefont {Ye}},\ }\href@noop {} {\bibfield  {journal} {\bibinfo
  {journal} {Nature}\ }\textbf {\bibinfo {volume} {542}},\ \bibinfo {pages}
  {66} (\bibinfo {year} {2017})}\BibitemShut {NoStop}%
\bibitem [{\citenamefont {Goban}\ \emph {et~al.}(2018)\citenamefont {Goban},
  \citenamefont {Hutson}, \citenamefont {Marti}, \citenamefont {Campbell},
  \citenamefont {Perlin}, \citenamefont {Julienne}, \citenamefont {D'Incao},
  \citenamefont {Rey},\ and\ \citenamefont {Ye}}]{2018_Ye_Nature}%
  \BibitemOpen
  \bibfield  {author} {\bibinfo {author} {\bibfnamefont {A.}~\bibnamefont
  {Goban}}, \bibinfo {author} {\bibfnamefont {R.~B.}\ \bibnamefont {Hutson}},
  \bibinfo {author} {\bibfnamefont {G.~E.}\ \bibnamefont {Marti}}, \bibinfo
  {author} {\bibfnamefont {S.~L.}\ \bibnamefont {Campbell}}, \bibinfo {author}
  {\bibfnamefont {M.~A.}\ \bibnamefont {Perlin}}, \bibinfo {author}
  {\bibfnamefont {P.~S.}\ \bibnamefont {Julienne}}, \bibinfo {author}
  {\bibfnamefont {J.~P.}\ \bibnamefont {D'Incao}}, \bibinfo {author}
  {\bibfnamefont {A.~M.}\ \bibnamefont {Rey}}, \ and\ \bibinfo {author}
  {\bibfnamefont {J.}~\bibnamefont {Ye}},\ }\href {\doibase
  10.1038/s41586-018-0661-6} {\bibfield  {journal} {\bibinfo  {journal}
  {Nature}\ }\textbf {\bibinfo {volume} {563}},\ \bibinfo {pages} {369}
  (\bibinfo {year} {2018})}\BibitemShut {NoStop}%
\bibitem [{\citenamefont {Gra\ss{}}(2019)}]{2019_Grass_PRL}%
  \BibitemOpen
  \bibfield  {author} {\bibinfo {author} {\bibfnamefont {T.}~\bibnamefont
  {Gra\ss{}}},\ }\href {\doibase 10.1103/PhysRevLett.123.120501} {\bibfield
  {journal} {\bibinfo  {journal} {Phys. Rev. Lett.}\ }\textbf {\bibinfo
  {volume} {123}},\ \bibinfo {pages} {120501} (\bibinfo {year}
  {2019})}\BibitemShut {NoStop}%
\bibitem [{\citenamefont {Glaetzle}\ \emph {et~al.}(2014)\citenamefont
  {Glaetzle}, \citenamefont {Dalmonte}, \citenamefont {Nath}, \citenamefont
  {Rousochatzakis}, \citenamefont {Moessner},\ and\ \citenamefont
  {Zoller}}]{glaetzle2014quantum}%
  \BibitemOpen
  \bibfield  {author} {\bibinfo {author} {\bibfnamefont {A.~W.}\ \bibnamefont
  {Glaetzle}}, \bibinfo {author} {\bibfnamefont {M.}~\bibnamefont {Dalmonte}},
  \bibinfo {author} {\bibfnamefont {R.}~\bibnamefont {Nath}}, \bibinfo {author}
  {\bibfnamefont {I.}~\bibnamefont {Rousochatzakis}}, \bibinfo {author}
  {\bibfnamefont {R.}~\bibnamefont {Moessner}}, \ and\ \bibinfo {author}
  {\bibfnamefont {P.}~\bibnamefont {Zoller}},\ }\href {\doibase
  10.1103/PhysRevX.4.041037} {\bibfield  {journal} {\bibinfo  {journal} {Phys.
  Rev. X}\ }\textbf {\bibinfo {volume} {4}},\ \bibinfo {pages} {041037}
  (\bibinfo {year} {2014})}\BibitemShut {NoStop}%
\bibitem [{\citenamefont {Weber}\ \emph {et~al.}(2017)\citenamefont {Weber},
  \citenamefont {Tresp}, \citenamefont {Menke}, \citenamefont {Urvoy},
  \citenamefont {Firstenberg}, \citenamefont {Büchler},\ and\ \citenamefont
  {Hofferberth}}]{weber2017calculation}%
  \BibitemOpen
  \bibfield  {author} {\bibinfo {author} {\bibfnamefont {S.}~\bibnamefont
  {Weber}}, \bibinfo {author} {\bibfnamefont {C.}~\bibnamefont {Tresp}},
  \bibinfo {author} {\bibfnamefont {H.}~\bibnamefont {Menke}}, \bibinfo
  {author} {\bibfnamefont {A.}~\bibnamefont {Urvoy}}, \bibinfo {author}
  {\bibfnamefont {O.}~\bibnamefont {Firstenberg}}, \bibinfo {author}
  {\bibfnamefont {H.~P.}\ \bibnamefont {Büchler}}, \ and\ \bibinfo {author}
  {\bibfnamefont {S.}~\bibnamefont {Hofferberth}},\ }\href {\doibase
  10.1088/1361-6455/aa743a} {\bibfield  {journal} {\bibinfo  {journal} {Journal
  of Physics B: Atomic, Molecular and Optical Physics}\ }\textbf {\bibinfo
  {volume} {50}},\ \bibinfo {pages} {133001} (\bibinfo {year}
  {2017})}\BibitemShut {NoStop}%
\bibitem [{\citenamefont {Anderegg}\ \emph {et~al.}(2019)\citenamefont
  {Anderegg}, \citenamefont {Cheuk}, \citenamefont {Bao}, \citenamefont
  {Burchesky}, \citenamefont {Ketterle}, \citenamefont {Ni},\ and\
  \citenamefont {Doyle}}]{Anderegg2019AnOT}%
  \BibitemOpen
  \bibfield  {author} {\bibinfo {author} {\bibfnamefont {L.}~\bibnamefont
  {Anderegg}}, \bibinfo {author} {\bibfnamefont {L.~W.}\ \bibnamefont {Cheuk}},
  \bibinfo {author} {\bibfnamefont {Y.}~\bibnamefont {Bao}}, \bibinfo {author}
  {\bibfnamefont {S.}~\bibnamefont {Burchesky}}, \bibinfo {author}
  {\bibfnamefont {W.}~\bibnamefont {Ketterle}}, \bibinfo {author}
  {\bibfnamefont {K.-K.}\ \bibnamefont {Ni}}, \ and\ \bibinfo {author}
  {\bibfnamefont {J.~M.}\ \bibnamefont {Doyle}},\ }\href {\doibase
  10.1126/science.aax1265} {\bibfield  {journal} {\bibinfo  {journal}
  {Science}\ }\textbf {\bibinfo {volume} {365}},\ \bibinfo {pages} {1156}
  (\bibinfo {year} {2019})}\BibitemShut {NoStop}%
\bibitem [{\citenamefont {Yu}\ \emph {et~al.}(2019)\citenamefont {Yu},
  \citenamefont {Cheuk}, \citenamefont {Kozyryev},\ and\ \citenamefont
  {Doyle}}]{Yu_2019}%
  \BibitemOpen
  \bibfield  {author} {\bibinfo {author} {\bibfnamefont {P.}~\bibnamefont
  {Yu}}, \bibinfo {author} {\bibfnamefont {L.~W.}\ \bibnamefont {Cheuk}},
  \bibinfo {author} {\bibfnamefont {I.}~\bibnamefont {Kozyryev}}, \ and\
  \bibinfo {author} {\bibfnamefont {J.~M.}\ \bibnamefont {Doyle}},\ }\href
  {\doibase 10.1088/1367-2630/ab428d} {\bibfield  {journal} {\bibinfo
  {journal} {New J. Phys.}\ }\textbf {\bibinfo {volume} {21}},\ \bibinfo
  {pages} {093049} (\bibinfo {year} {2019})}\BibitemShut {NoStop}%
\end{thebibliography}%
\bibliographystyle{apsrev4-1}

\end{document}